\documentclass[12pt]{iopart}
\pdfoutput=1
\usepackage{iopams}
\usepackage{graphicx}
\usepackage{citesort} 
\usepackage{subfig}
\usepackage{color}
\begin{document}

\title[]{Investigation of the effectiveness of non-inductive `multi-harmonic' electron cyclotron current drive through modeling multi-pass absorptions in the EXL-50 spherical tokamak}


\author{Debabrata Banerjee, Shaodong Song, Huasheng Xie, Bing Liu, Mingyuan Wang, Wenjun Liu, Bin Chen, Lei Han, Di Luo, Yunyang Song, Xianming Song, Minsheng Liu, Yuejiang Shi, Y. K. Martin Peng and the EXL-50 team}

\address{Hebei Key Laboratory of Compact Fusion, Langfang 065001, China}
\address{ENN Science and Technology Development Co. Ltd., Langfang 065001, China}

\ead{\mailto{banerjeed@enn.cn}}
\ead{\mailto{shiyuejiang@enn.cn}}

\author{Yu. V. Petrov and R. W. Harvey}

\address{CompX, Del Mar, CA, USA}


\begin{abstract}
	The effectiveness of multiple electron cyclotron resonance (ECR) harmonics has been thoroughly investigated in context of high current drive efficiency, generally observed in fully non-inductive operation of the low aspect ratio EXL-50 spherical tokamak (ST) powered by electron cyclotron (EC) waves. The Fokker-Plank equation is numerically solved to obtain electron distribution function, under steady state of the relativistic nonlinear Coulomb collision and quasi-linear diffusion operators, for calculating plasma current driven by the injected EC wave. For the extra-ordinary EC wave, simulation results unfold a mechanism by which electrons moving around the cold second harmonic ECR layer strongly resonate with higher harmonics via the relativistic Doppler shifted resonance condition. This feature is in fact evident above a certain value of input EC wave power in simulation, indicating it to be a non-linear phenomenon. Similar to the experimental observation, high efficiency in current drive (over 1 A/W) has indeed been found in simulation for a typical low density ($\sim 1\times10^{18}~m^{-3}$), low temperature ($\lesssim 100$ eV) plasma of EXL-50 by taking into account multi-pass absorptions in our simulation model. However, such characteristic is not found in the ordinary EC-wave study for both single-pass and multi-pass simulations, suggesting it as inefficient in driving current on our ST device. 
\end{abstract}
%
%
\submitto{\NF}
%
%
%
\section{Introduction}
The interaction of radio frequency (RF) wave's electro-magnetic field with gyrating electrons in plasma has long been an interesting and important topic of study in perspective of electron cyclotron resonance (ECR) phenomena, governing plasma heating~\cite{prater03,karney78} and current drive in a localized region of fusion plasma~\cite{luce99,harvey97}, acceleration of electrons~\cite{menyuk88}, and the control of magneto hydrodynamic instability~\cite{harvey01}. There are circumstances under intense electric field of a RF wave, when an electron gets opportunity to resonate with multiple harmonics simultaneously to gain much high energy from the wave field, and its motion becomes stochastic in nature~\cite{menyuk88}. The generation of energetic electrons by ECR phenomena had also been verified by laboratory experiments in which the 2nd harmonic was found to be more effective than the fundamental ECR~\cite{ikegami73}. The possibility of non-inductive start-up of a hot plasma current in future large fusion reactors by means of electron cyclotron resonance heating (ECRH) is theoretically investigated recently~\cite{maekawa18a,maekawa18b}. An electron, moving with velocity $v_{\parallel}$ along an ambient magnetic field of value $B$, resonates with the electric field of EC wave propagating with a wave number $k_{\parallel}$,  parallel to that magnetic field, via the condition $\omega - k_{\parallel}v_{\parallel} = n\Omega/\gamma$; where $\gamma$ is the relativistic factor, n stands for the different ECR harmonics and $\Omega$ defines the non-relativistic electron cyclotron frequency~\cite{erckmann94}.  Usually, in all conventional tokamaks of large aspect ratio, one or at most two ECR harmonics co-exist inside vacuum vessel. This scenario changes for a new generation low aspect ratio ST which, under purposeful matching of toroidal magnetic field profile and EC wave frequency, contains several ECR harmonics. Such a device named EXL-50, built and owned by the ENN (Energy iNNovation) science and technology development private limited company stationed at Langfang in China, starting its high power operation in the beginning of year 2020 until now, has routinely achieved high current ($I_p$) drive efficiency (over 1 A/W) in fully non-inductive plasma discharges solely powered by ECRH~\cite{shi21,ishida21}. There are five number of cold plasma ECR harmonics present inside its vacuum vessel if the toroidal field (TF) coil current is set close to 100 kA. Similarly, high efficiency in plasma current was reported in last two decades on other low aspect ratio devices as well; for example, the LATE~\cite{maekawa05} and QUEST~\cite{idei17} STs observed high current in ECRH driven fully non-inductive operations (around 75 kA current with about 113 kW injected RF power in QUEST~\cite{onchi21}), the MAST device~\cite{shevchenko10} achieved significant amount of current (as high as 73 kA current with upto 60 kW of injected RF power~\cite{shevchenko15}) in electron Bernstein waves driven plasma operation.

In all of these non-inductive ECRH experiments on ST devices, the hard x-ray measurement supports that plasma current is mainly carried by a group of high velocity electrons, known as energetic electrons. However, a complete understanding of the physical mechanism that drives electron to be so energetic and its apparently good confinement in carrying big amount of current remains unachieved. Our effort in the present article is to address this point in light of the effectiveness of multi-harmonic ECRs. A numerical study has been carried out using a Fokker-Plank (FP) code CQL3D~\cite{harvey92} in conjunction with a ray-tracing code GENRAY~\cite{harvey94} to simulate $I_p$ for a typical EXL-50 discharge. The extra-ordinary EC wave (abbreviated as X-wave or X-mode) is found to be much more efficient than the ordinary EC wave (O-wave or O-mode) in terms of power deposition and current drive in single-pass simulation. In our study, it is revealed that electrons resonate with multiple harmonics of the X-wave ECR to become highly energetic, and as a result, a significant amount of current is driven. This fact also prevails for muti-pass X-wave simulations as well to generate $I_p$ value in the same order of experimentally measured value. However, the O-mode EC wave simulation has output negligible amount of current even in multi-pass simulation. The role of higher harmonic ECRs for resonating with high energy tail of electron distribution and driving consequently large amount of plasma current becomes more evident in simulation when we step the input power up. This fact is identified to have a probable connection with the quadratic dependence of wave induced quasi-linear (QL) diffusion coefficients on the injected EC wave's electric field. These all results will be presented in detail in the following sections with in-depth analysis and appropriate discussion.      
\section{EXL-50 plasma powered by electron cyclotron range of frequency (ECRF) wave}
\begin{figure}[htbp]
\subfloat[]{\includegraphics[height=8.5cm,width=7.5cm]{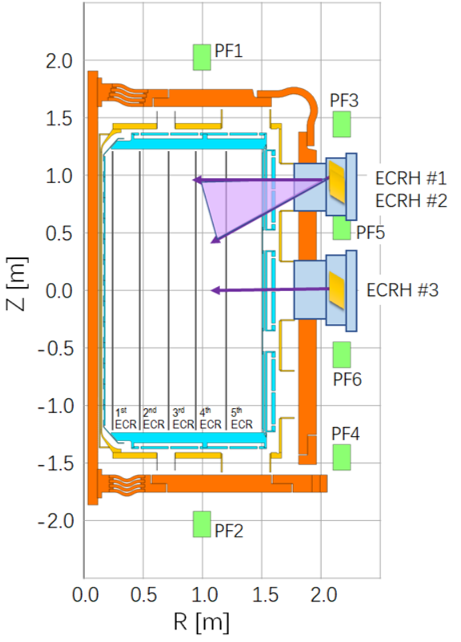}}
\subfloat[]{\includegraphics[height=8.5cm,width=8.5cm]{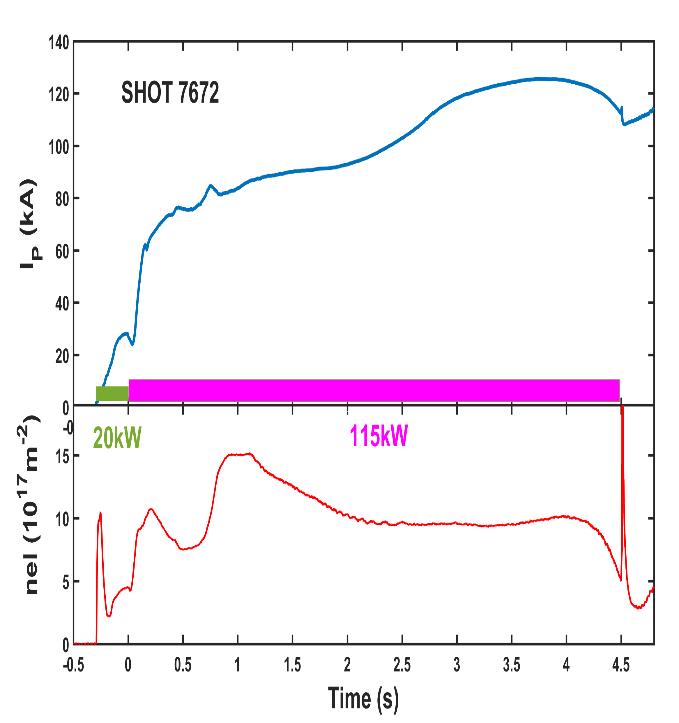}}%
	\caption{\label{fig1exper}(a) Schematic diagram of the EXL50 device in poloidal cross-section showing the ports of ECRH sources (in upper and middle planes), locations of six PF coils and the cold ECR harmonic layers as solid black lines. (b) The waveforms of time evolution for one high current drive efficiency discharge $\#7672$. One gyrotron (ECRH1) injected 20 kW from -0.3 s to 0 s and the other gyrotron (ECRH2) injected 115 kW in (0 - 4.5)s duration. $I_P$ and $nel$ are the experimentally measured plasma current and line integrated electron density respectively.}

\end{figure}
The non-solenoidal spherical torus EXL-50 has started operation nearly two years ago. Having envision to explore fully non-inductive long pulse steady state plasma operation, as first step, it utilizes the power of ECRF wave from multiple gyrotrons to startup, heat plasma and subsequently drive current~\cite{shi21}. In a schematic diagram drawn in poloidal cross-section, figure~\ref{fig1exper}(a) shows locations of the six poloidal field coils and three ECRH sources of 28 GHz frequency each - one installed in an equatorial port, and others in an upper mid-plane port being 30 degrees apart from the equatorial one in toroidal direction. A narrow centre post cylinder of radius $\sim$ 17 cm is installed which embraces inner legs of the twelve single turn TF coils capable of conducting upto 100 kA current, but without any solenoidal coils to induce toroidal loop voltage. The hot plasma is kept detached from the device wall with the help of a set of limiters made of tungsten metal installed on the centre post and outboard vessel wall. Also drawn in this diagram are layers of the fundamental and other harmonics (2nd, 3rd, 4th \& 5th) of cold plasma ECR present inside the cylindrical shaped vacuum vessel. 
In the discharge $\#7672$ shown in figure~(1b), the closed field lines plasma was found to extend from the fundamental to 4th ECR harmonic, with the 5th one being in open field line plasma region. This discharge is chosen from an experimental campaign on EXL-50 in the last quarter of 2021, when one high power gyrotron (ECRH2) was utilized to inject power (100 - 120) kW, and one low power gyrotron (ECRH1) was used to inject (10 - 20) kW; here the power is measured in the matching optical unit (MOU) that is close to output power from gyrotron. As seen in figure~(1b), $I_p$ exhibits a jump after ECRH2 was turned on, then it steadily increases with time to reaching a level more than 1 A/W efficiency. In fact, high plasma current has been routinely observed ranging from 50 kA to 150 kA with the input ECRH power varied from 20 kW upto 120$~$kW in fully ECRH driven discharges~\cite{shi21}. Line density was measured by a single-chord tangential microwave interferrometer~\cite{li21} with its value recorded nearly $1\times10^{18}~m^{-2}$ in the flat-top until a sudden peak appeared after the source ECRH2 had been turned off. 

ECRH led breakdown and heating of plasma cause to generate a large number
of  energetic electrons measured by a hard X-ray diagnostic system~\cite{shi21,cheng21}. In contrary to background highly populated thermal electrons ($<100$ eV) mostly being confined within the closed field line plasma, energetic electrons are speculated to roam around in both closed and open field lines plasma regions to carry major portion of plasma current. In full orbit simulations, trajectories of such electrons in the magnetic field structure of these discharges are studied to find that even electrons upto several MeV can be confined inside vacuum vessel. Such dominancy of energetic electrons over thermal electrons was also reported in similar type of experiment performed on LATE~\cite{uchida10} and QUEST \cite{idei17}. Based on experimental finding, one's intuition could be that the thermal and energetic population of electrons perhaps behave as two separate fluids identity of electrons to maintain momentum balance with ions. This ideology has been implemented in a multi-fluid equilibrium code to reconstruct flux information, and thereby estimate plasma current in force balance~\cite{ishida21}. Such calculation results in a broadened current density profile of energetic electrons in plasma covering both closed and open field line regions as shown in figure~8(b) of reference~\cite{shi21}. 
In a low density, low temperature plasma at which currently EXL-50 is operating, wave power deposition on electrons during first pass is within $10\%$ of the injected power for X-wave, and further one order less for the O-wave, as found in numerical simulation (shown in the next section). Therefore, in the experiment, we may reasonably consider that an EC beam, launched  primarily with the O-mode type polarisation from antenna, passes through the plasma many times upon constantly being reflected on the vessel wall made of stainless steel and on limiter surfaces. 
 In reality, the O-mode EC wave has certain probability to convert into other X-mode EC wave upon reflection on a shiny wall with low reflection loss~\cite{maekawa12}. Therefore, from statistical viewpoint, many times reflected wave may be modeled as having nearly equal probability of being O- and X-wave while it crosses plasma closed flux region many times. Based on this idea, the present article demonstrates multi-pass simulation results considering half of absorbed power attributed to the X-wave, and other half to the O-wave out of the total injected power. The scenario of multi-pass absorptions in the EXL-50 plasma has actually made it quite different from the ECRF wave heated high temperature plasmas in conventional toroidal devices. The generation of large number energetic electrons and their instrumental role towards high current drive efficiency is thought as an outcome of this. 
\section{Numerical study of the role of multi-harmonic resonances on single-pass absorption }
In order to understand the advantage of having multiple number of ECR harmonics co-existed inside EXL-50 plasma over only a single resonance in terms of current drive efficiency, a numerical simulation study has been carried out employing a widely used code CQL3D coupled with another code GENRAY. The CQL3D code~\cite{harvey92,kerbel85} has capability to evolve electron distribution function by numerically solving the Fokker-Plank equation in 2D momentum space considering the Coulomb collision and QL diffusion operators as contributing terms. The solution space is set up onto a 2D co-ordinates comprising of momentum along the magnetic field and gyro-averaged momentum across the field. Relativistic Coulomb collisions among thermal and high velocity tail electrons, and electrons with Maxwellian ions are modeled by implementing the Braams-Karney relativistic nonlinear potential functions~\cite{braams89}. Wave-electron interaction is modeled by the well-accepted  quasi-linear diffusion theory. 
To calculate QL diffusion coefficients, all necessary information such as, ray element position, power along rays, electric field polarisation and local magnetic field value, has been generated by ray-tracing simulation in GENRAY~\cite{harvey94}, and then passed on to CQL3D as input. The bounce averaged FP equation solved in CQL3D for our simulation purpose is expressed as,
\begin{eqnarray}
	\lambda \frac{\partial}{\partial t}f_0 \left(u_0,\theta_0,\rho,t \right) = C(f_0) + Q(f_0)
\end{eqnarray}
where $f_0$ is the electron distribution as a function of the momentum-per-mass ($u_0$) and pitch angle ($\theta_0$) at the minimum-B point on each flux surface, $\rho$ the generalized radial coordinate, and t the time. Jacobian $ \lambda =  v_{\parallel 0} \tau_b$ accounts for the volume of the flux surface occupied per perpendicular area at the midplane, here $v_{\parallel 0}$ stands for parallel speed, and $\tau_b$ the bounce (or transit) period. The operator C calculates bounce averaged relativistic nonlinear Coulomb collision effect~\cite{braams89}, and Q is the full bounce averaged relativistic Stix QL operator~\cite{stix}; full mathematical form of these operators can be found in the CQL3D manual~\cite{cql3d}.
The initial distribution of electrons is considered as Maxwellian which evolves in time to become non-Maxwellian under the play between collision and diffusion operators. All ion species are modeled as background Maxwellian distributions to be engaged in collision with the evolving non-Maxwellian distribution of electrons.  The coupled code suite GENRAY-CQL3D was validated successfully by many earlier research studies to predict and explain EC wave heating and current drive experiment on several devices worldwide~\cite{prater08,harvey02,zheng18,petty02}. 
\subsection{Equilibrium setup}
\begin{figure}[htbp]
~~~~~~~~~~~~\subfloat[]{\includegraphics[height=7.5cm,width=5.5cm]{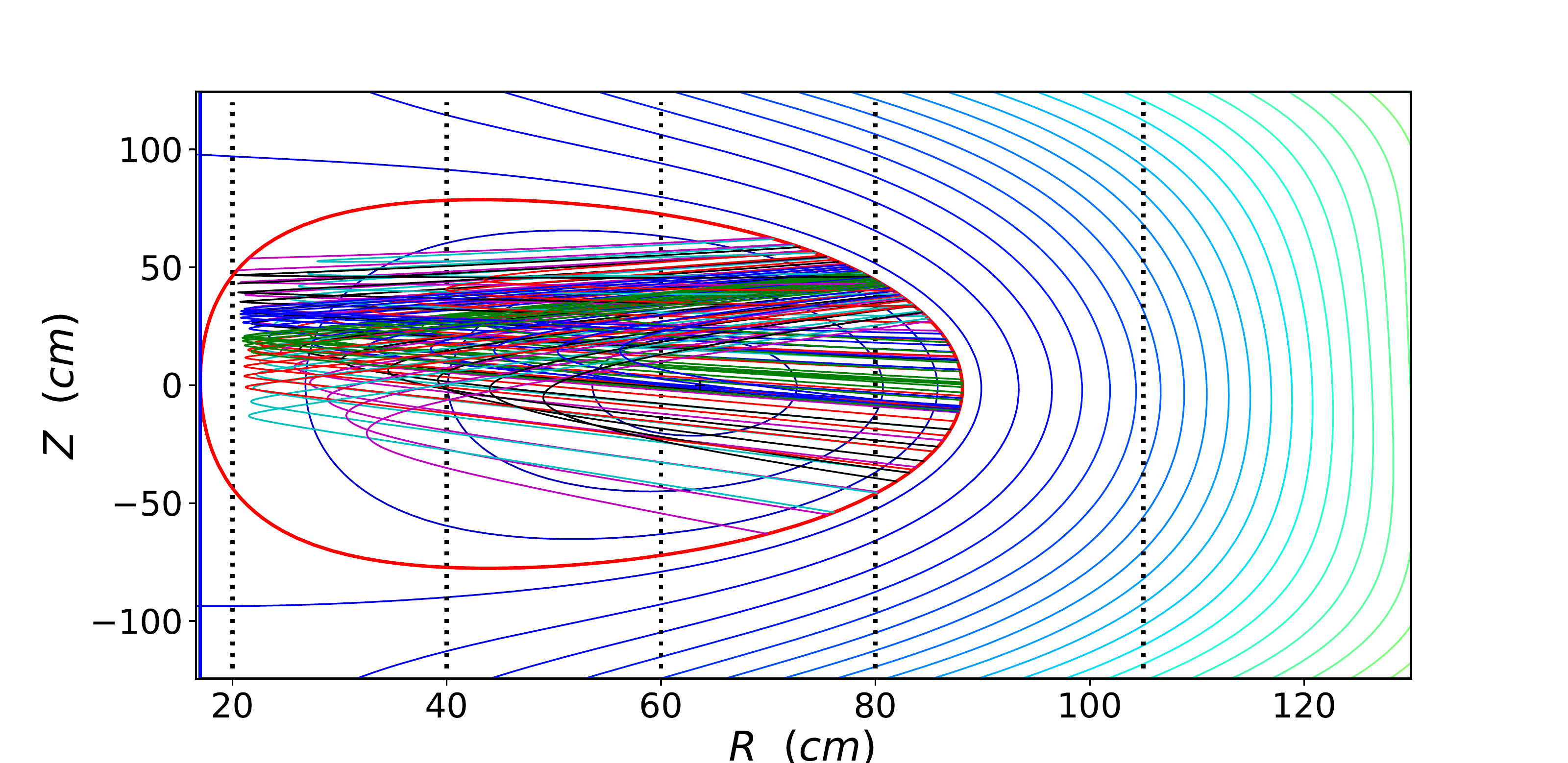}}
~~\subfloat[]{\includegraphics[height=7.5cm,width=6.5cm]{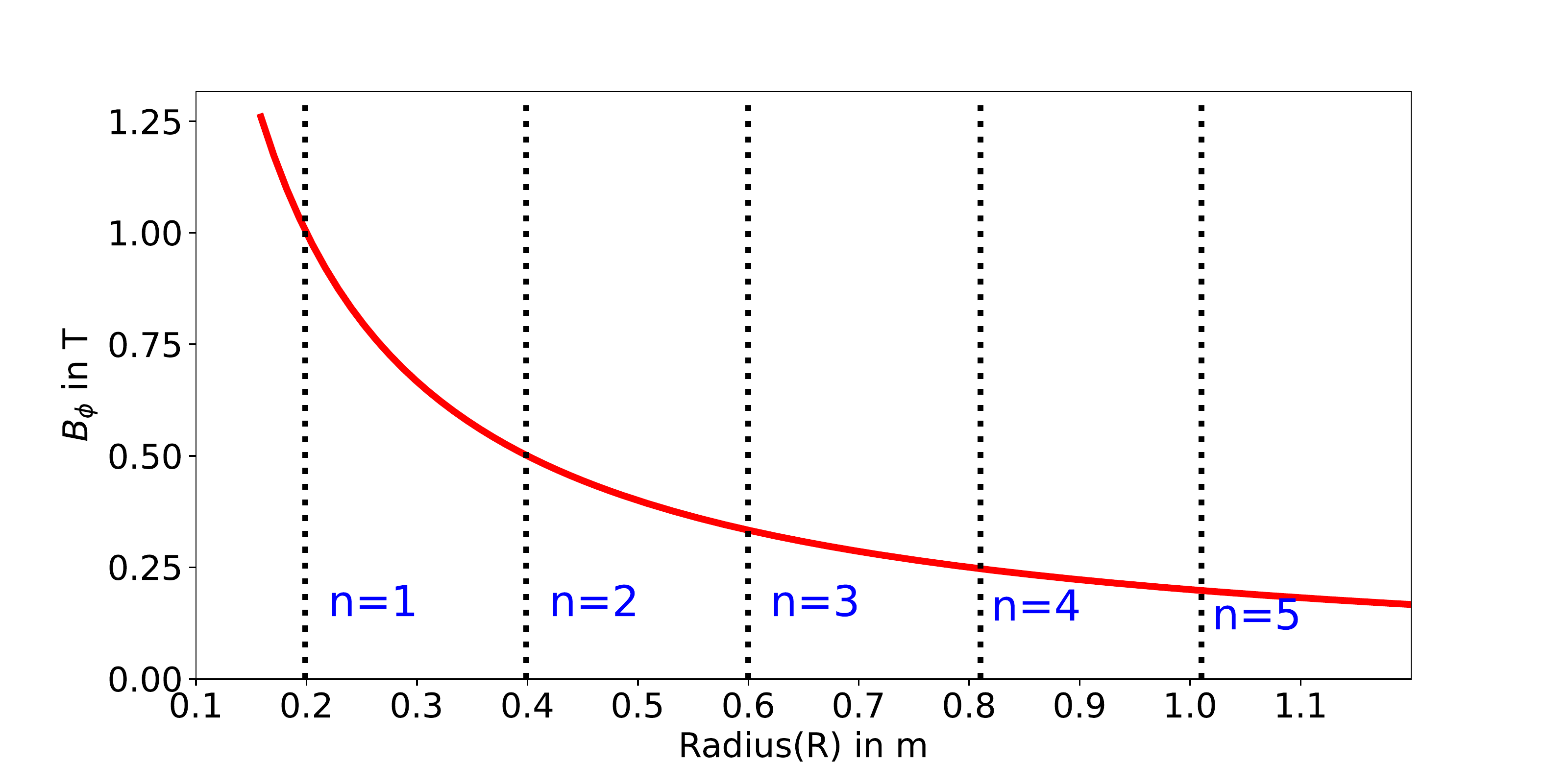}}\\
.\subfloat[]{\includegraphics[height=4.5cm,width=15.5cm]{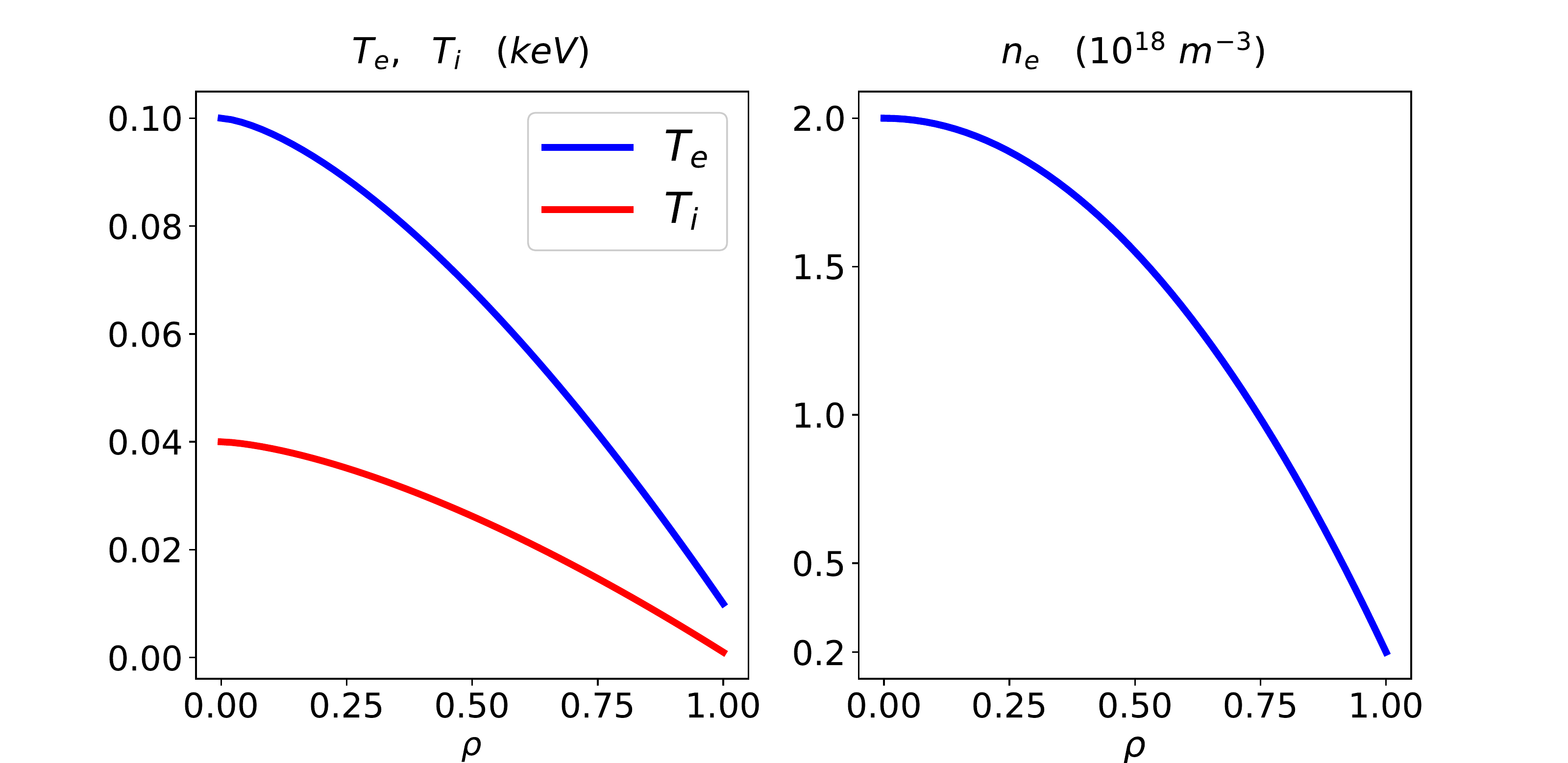}}	
	\caption{\label{fig2eqx} (a) Contour plot of poloidal flux with illustration of the X-mode ray-trajectory and cold plasma ECR harmonic layers. (b) toroidal magnetic field vs. major radius with the cold ECR harmonic locations. (c) Assumed profiles of species density and temperature following equation (2), core electron density $n_{oe}=2\times 10^{18}~m^{-3}$, core electron temperature $T_{oe}=100$ eV, core ion temperature $T_{oi}=40 $ eV. For $Z_{eff}=1$,~ $n_{oe}=n_{oi}$. For $Z_{eff} > 1$, density of hydrogen ion and other impurity ions are adjusted so that all ions' cumulative density equals electron density maintaining quasi-neutrality, and temperatures of impurity ions are kept same with that of H-ion.}
\end{figure}
The equilibrium is reconstructed based on the PF and TF coil currents, and the plasma current of discharge $\#7672$ to give input magnetic flux profile to GENRAY and CQL3D. As experimentally measured data for density and temperature radial profiles are not available at present, we use model algebraic profiles defined by,
\begin{eqnarray}
	n_{s} = \left(n_{os}-n_{bs} \right)\left(1-\rho^{\alpha} \right)^{\beta} + n_{bs}~;~~~
	T_{s} = \left(T_{os}-T_{bs} \right)\left(1-\rho^{\alpha} \right)^{\beta} + T_{bs}
\end{eqnarray}
Here, the radial co-ordinate is defined as $\rho=\sqrt{\Phi/\Phi_{lim}}$, $\Phi$ being the toroidal flux at respective flux surface, $\Phi_{lim}$ the same at last closed flux surface (LCFS). The density and temperature of different species s - meaning electron (e), hydrogen ion (i) and impurities - at the magnetic axis and LCFS are termed as 
$\left(n_{os},T_{os}\right)$ and $\left(n_{bs},T_{bs}\right)$ respectively.  
Also importantly, both GENRAY and CQL3D simulations have the same set-up of density, temperature, magnetic field and radial co-ordinate. The EC wave source is characterized in GENRAY by defining launching parameters similar to the experimental set-up i.e., the  upper-plane source ECRH2 launches 115 kW power at frequency 28 GHz directed at $15$ degrees poloidal angle relative to horizontal plane at the antenna location, and also obliquely in toroidal plane by an angle of 5 degrees with normal injection. Another important parameter to be defined in simulation is the injected beam width, which is kept sufficiently wide to match the relatively less focused beam at antenna.
\subsection{O-wave of 28GHz frequency }
\begin{figure}[htbp]
\subfloat[]{\includegraphics[height=6.5cm,width=8.5cm]{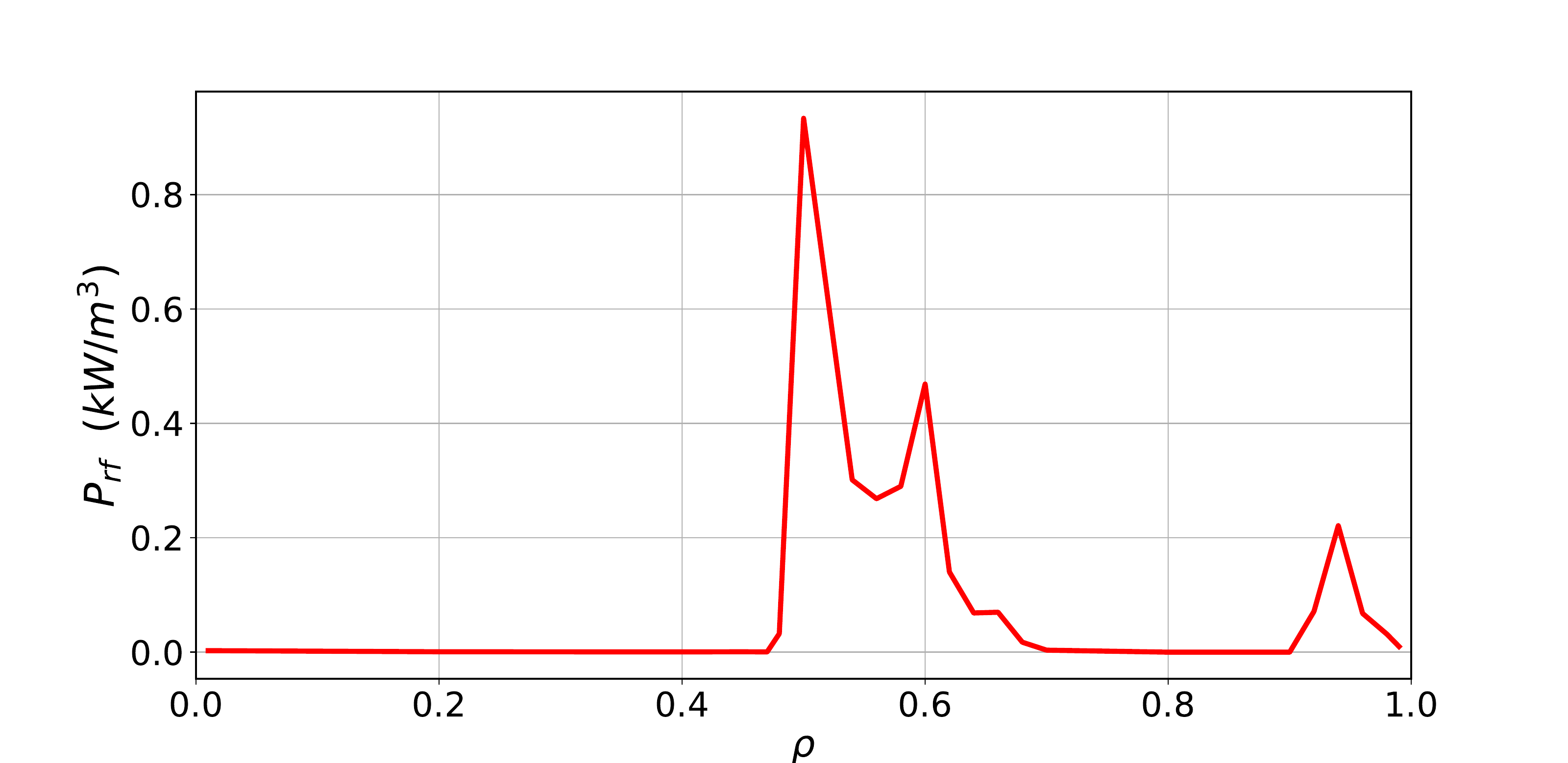}}
~~\subfloat[]{\includegraphics[height=6.5cm,width=8.5cm]{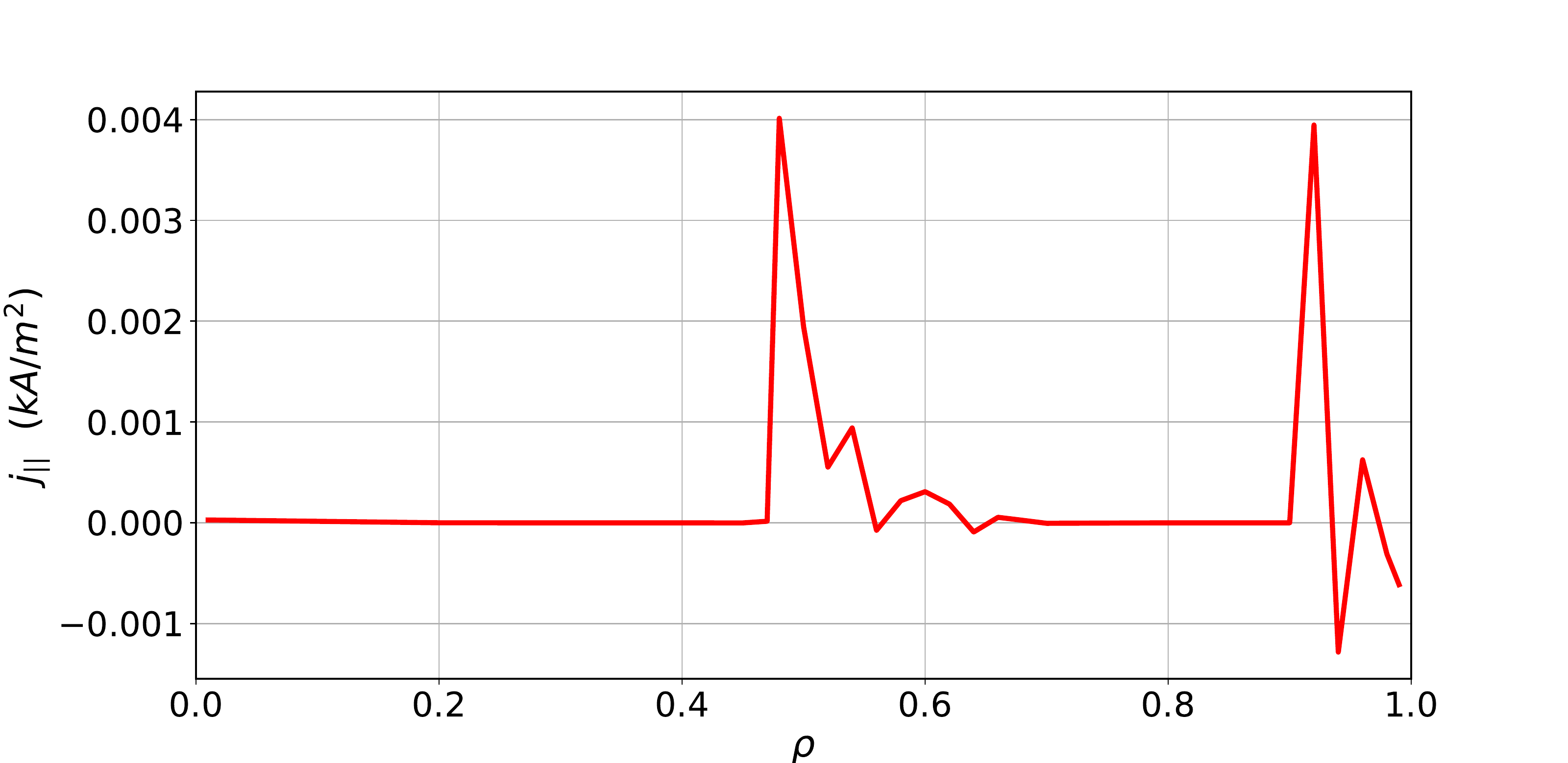}}%
	\caption{\label{fig3profO} O-wave power deposition profile (a), and current density profile (b) from CQL3D solution. Both profiles are localized around the fundamental and 2nd harmonic cold ECRs. }
\end{figure}

With this setup of EC ray-launch, and $n_s-T_s$ profiles as shown in figure~(2c), the GENRAY code was run first to perform ray-tracing to generate informations for calculating  QL diffusion coefficients. Next, the CQL3D code was run with the same density and temperature profiles, and the same equilibrium field setup to solve equation (1), and to calculate absorbed power and plasma current. As the polarisation of injected EC beam in experiment was primarily of O-wave type, we first have studied single pass absorption for O-mode root of the cold plasma dispersion relation in ray-tracing simulation. In CQL3D run, QL diffusion coefficients are calculated for total 5 harmonics ($n=1-5$) - from the fundamental ($n=1$) upto the 5th ECR harmonic ($n=5$) - in similarity with the number of cold ECR harmonics present inside EXL-50 as shown in figure~\ref{fig2eqx} (a,b). Absorbed power density profile with minor radius ($\rho$) in figure~\ref{fig3profO}(a) indicates absorption around the cold 2nd harmonic ECR at $\rho=0.5$ being stronger than that around the fundamental ECR ($n=1$) at $\rho=0.9$, even though damping for the fundamental ECR be stronger than on 2nd harmonic based on the ECRH theory prediction. This is because the $n=1$ cold resonance layer lies near to $\rho=1$ (LCFS) accessing much lower plasma temperature compared to the $n=2$ layer. On other hand, we know from ECRH theory that power deposition on $O2$ is at least one order lower than $X2$ for same density and temperature. 

\begin{figure}[htbp]
\subfloat[]{\includegraphics[height=6.5cm,width=9.5cm]{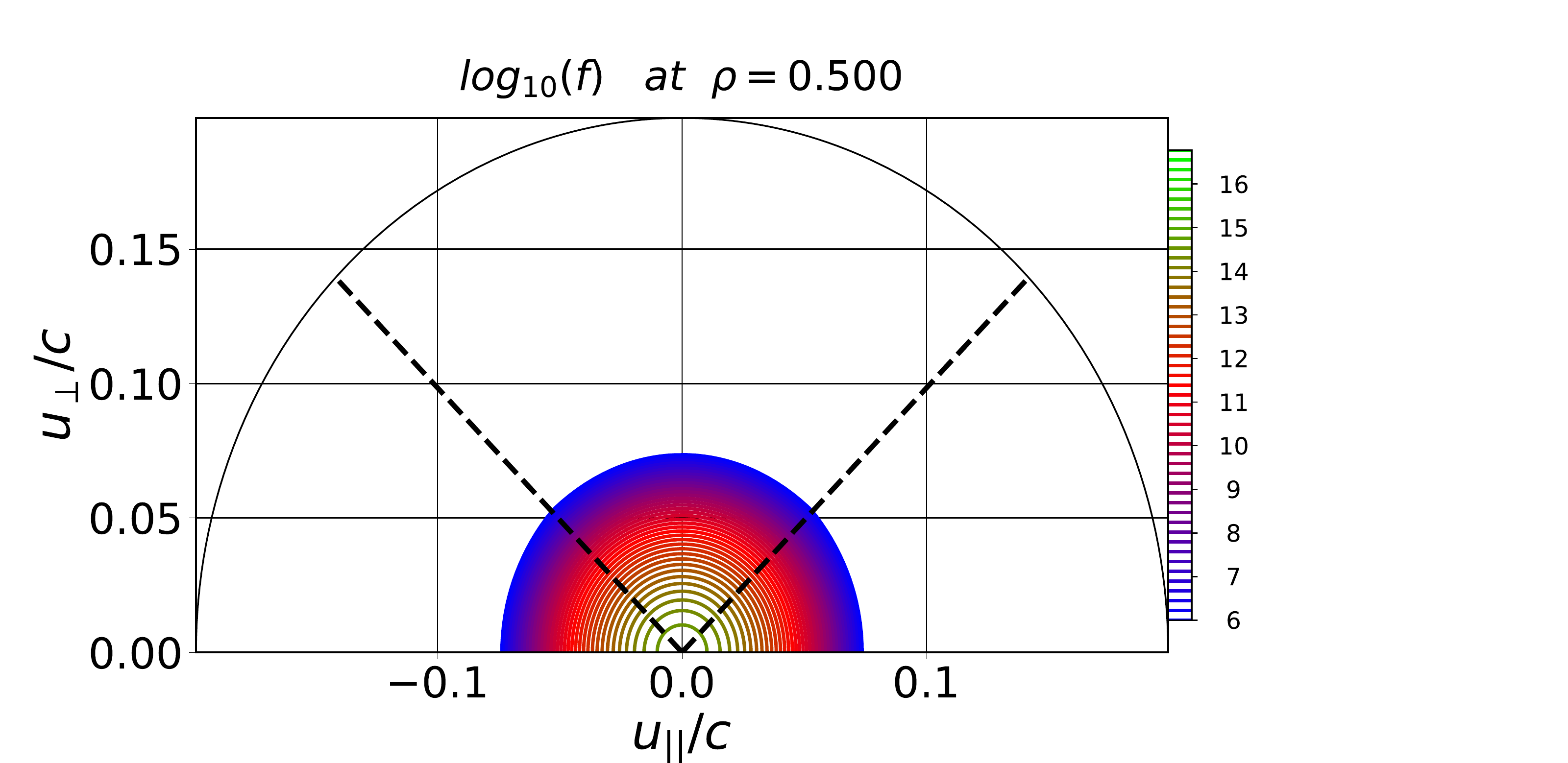}}
\subfloat[]{\includegraphics[height=7.5cm,width=7.5cm]{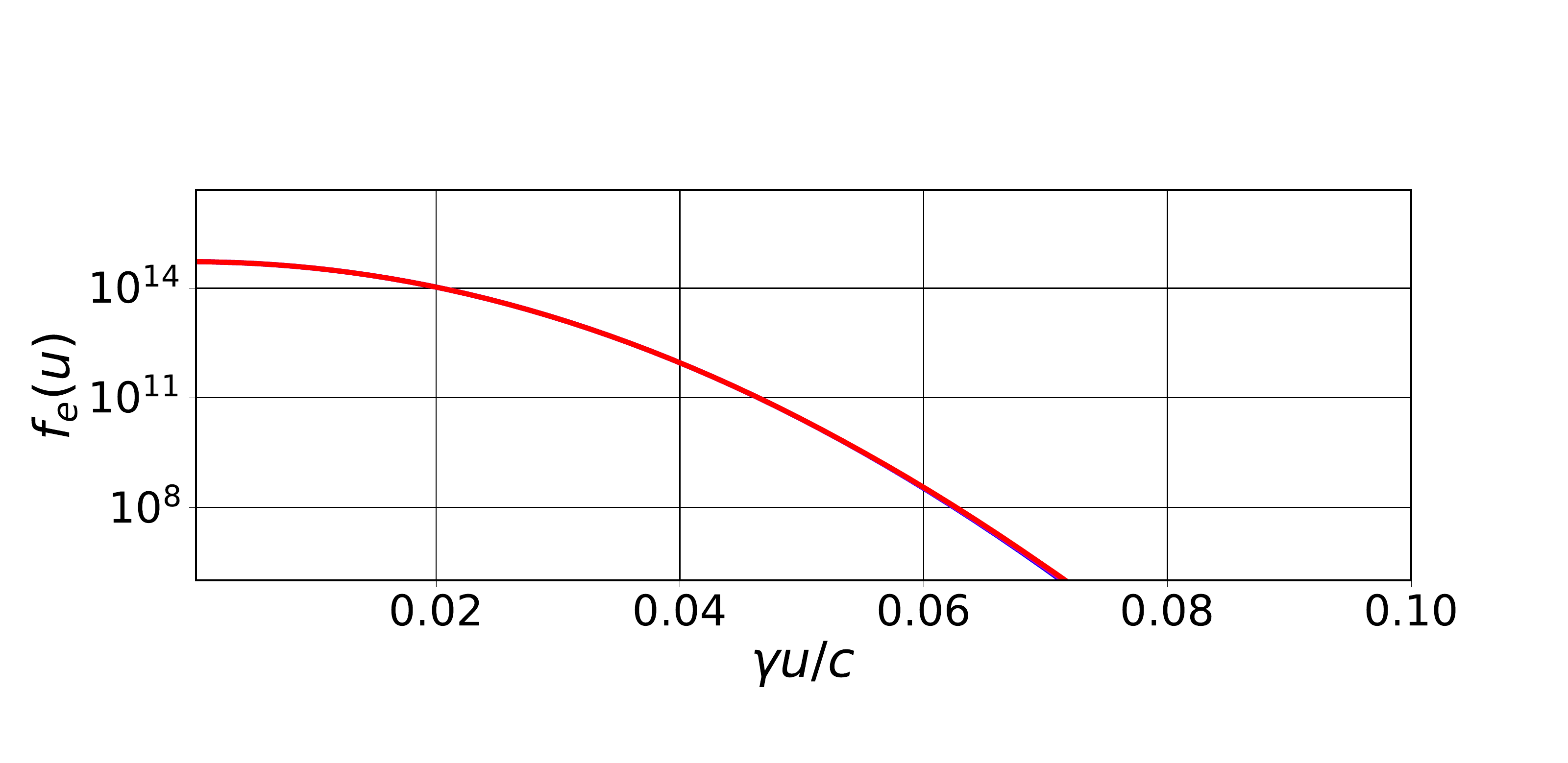}}%
	\caption{\label{fig4distO} (a) electron distribution function in relativistic ${\bf u}/c \equiv {\bf p}/mc$ space computed with CQL3D for the O-wave. Final solution remains almost equal to the starting Maxwellian form. (b) cuts through the 2D electron distribution function $f_e$ (of figure 4(a)) at four different constant pitch angles vs. $\gamma u /c$, $\gamma$ being the relativistic factor.}
\end{figure}
Therefore, a significantly lower amount of absorbed power (20 W on $O1$, 280 W on $O2$ and negligible amount at higher harmonics $n=3-5$) has come out from CQL3D run. As a result, driven plasma current at all harmonics is negligibly small (figure~\ref{fig3profO}(b)), projecting O-wave to be inefficient in driving an initial low density and low temperature plasma in our ST. This run amounts to a trivial current 0.2 A with basically no feature of multi-harmonic influence. This fact will be further confirmed by multi-pass simulation results in the next section. The electron distribution function ($f_e$), as it has emerged from the steady state solution of equation~(1), remains almost unchanged from the initial Maxwellian form (2D plot in figure~\ref{fig4distO}(a), and 1D cuts of $f_e$ at constant pitch angles in figure~\ref{fig4distO}(b)) without forming any energetic electron tail. Besides, we checked running simulations for different toroidal and poloidal launch angles but haven't found any improvement in current drive efficiency of the O-wave. This fact of the poor efficiency of O-wave in depositing power resembles with a recent ray-tracing simulation study on QUEST~\cite{onchi21}. Such finding has let us to investigate the efficacy of X-wave which may appear in present experiment via the mode conversion from O- to X-wave upon reflection on wall. 
\subsection{X-wave of 28GHz frequency}
Unlike the O-wave, similar type of numerical study with the X-wave has found high effectiveness of multi-harmonic EC resonances in current drive. Considering the same base parameters of magnetic field, density and temperature described in figure \ref{fig2eqx}, and equal wave launching set-up such as input power, frequency, injection angles and beam width used for the O-wave study, we have conducted ray-tracing in GENRAY for the X-mode branch of cold plasma dispersion, and then solved the FP equation in CQL3D. 
EXL-50 being operated with limiters and having inboard limiters neck to neck with the  cold plasma fundamental resonance layer, has high presence of impurities such as helium (He), carbon (C) etc. Keeping in mind that the resulting increased effective ion charge $Z_{eff}$ can alter  electron distribution and thus affect plasma current, we have decided to run simulations for two values of $Z_{eff}=2.5~ \&~ 4.0$. Although, a real value is not available from diagnostic measurements, this assumed value $Z_{eff}=4.0$ is expected to be close to the experimental one of discharge $\#7672$. 
\begin{figure}[htbp]
\subfloat[]{\includegraphics[height=7.0cm,width=8.5cm]{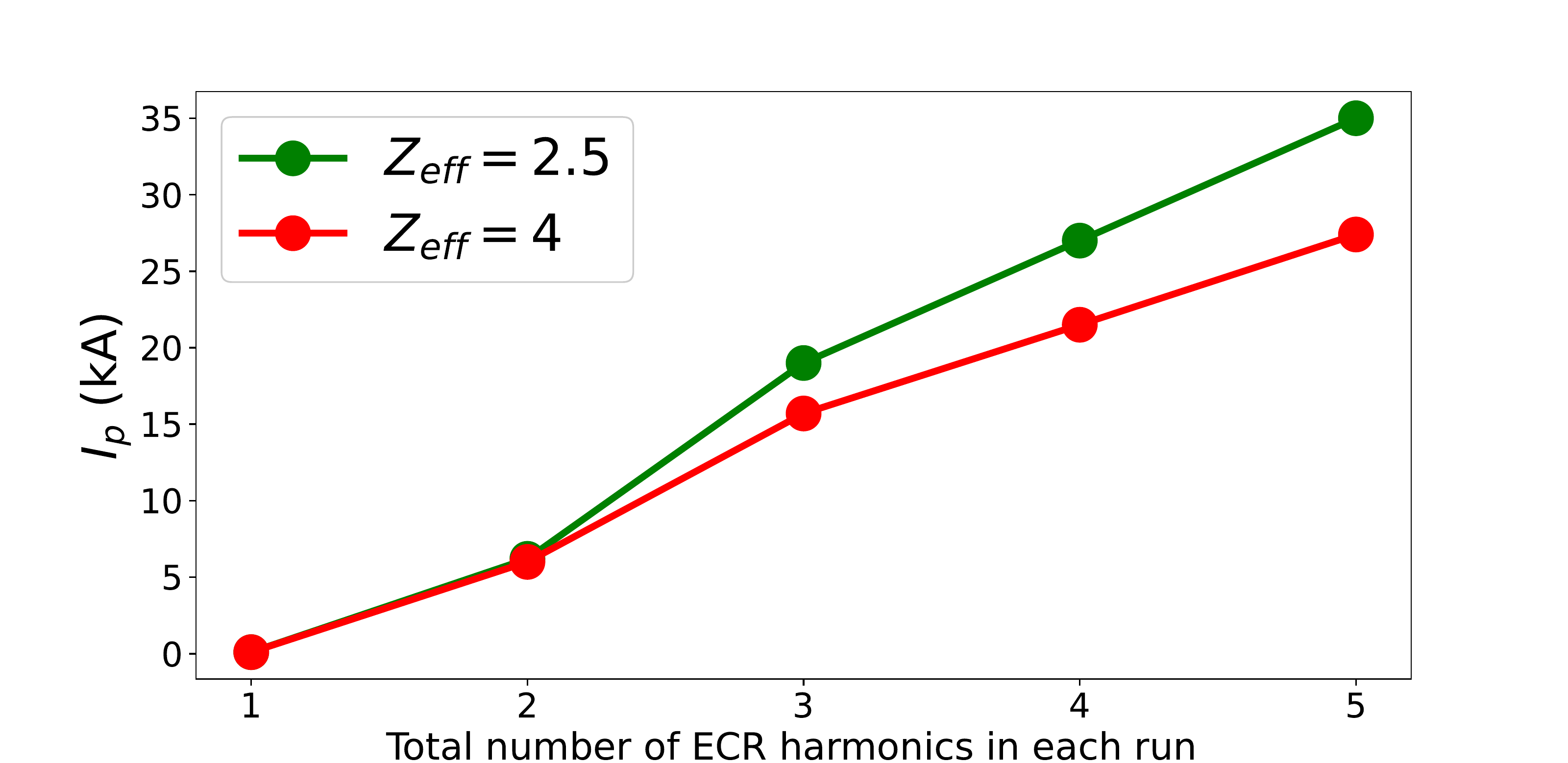}}
\subfloat[]{\includegraphics[height=7.0cm,width=9.0cm]{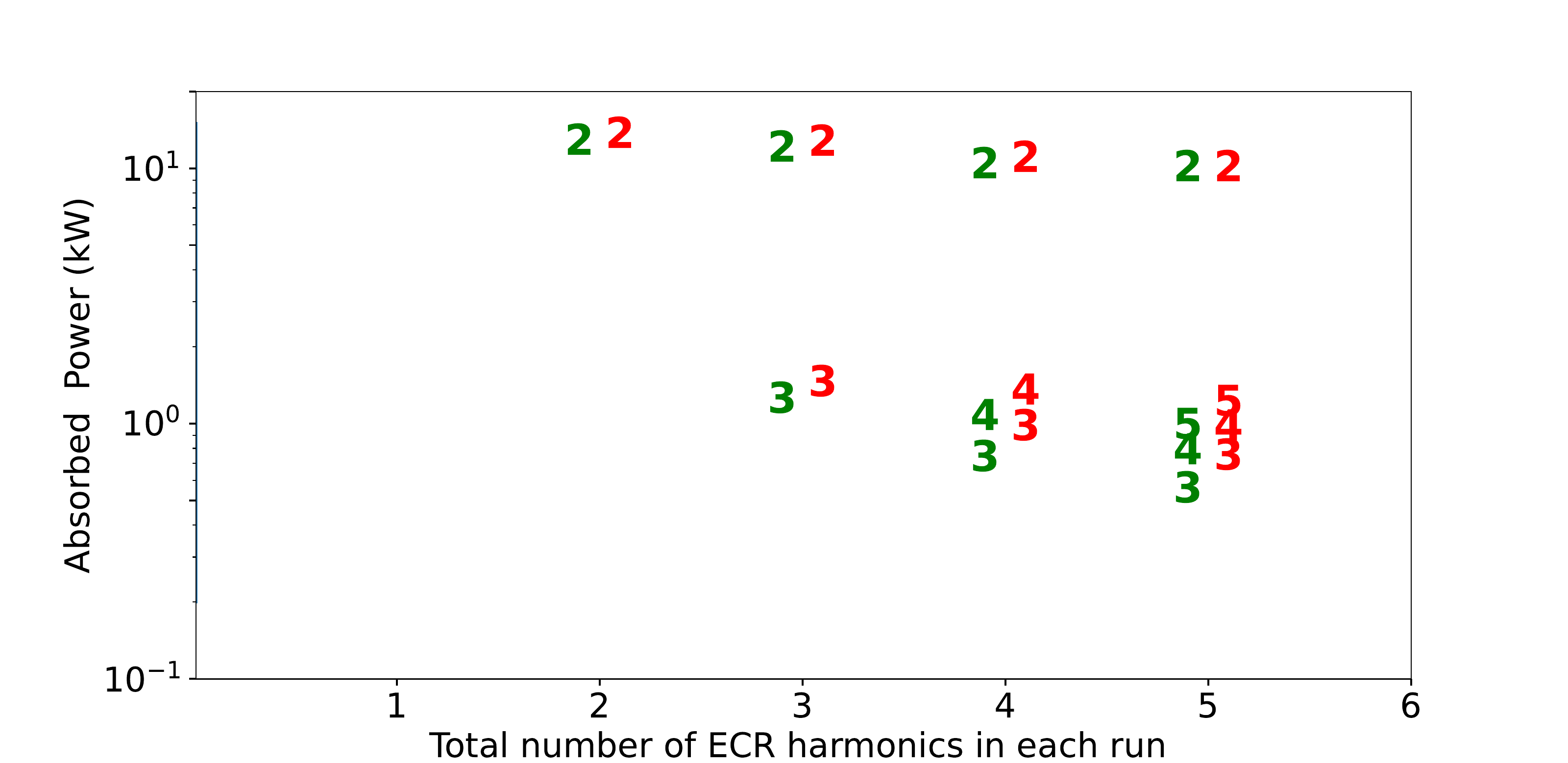}}%
	\caption{\label{fig5ipvsn} (a) The rise in plasma current with increasing number of ECR harmonics. The meaning of abscissa is described in subsection (3.3), (b) power sharing among harmonics for each run; mode numbers from each run are shown on graph at the corresponding absorbed power vertical location, green colour for $Z_{eff}=2.5$ and red for $Z_{eff}=4.0$}
\end{figure}
As shown in figure~(2a), all rays marked in different colours are reflected on the right hand cut-off frequency layer before reaching the fundamental resonance location as per the X-wave's cold plasma dispersion characteristic. To investigate the effectiveness of all harmonics in detail, we increase the number of harmonics step by step in CQL3D input following this order: 1st simulation run using only the fundamental harmonic, 2nd run with $n=1-2$ harmonics i.e. the fundamental and 2nd harmonics together, and so on upto the 5th run with $n=1-5$ harmonics. Toroidal plasma current obtained from CQL3D in every run is plotted against the respective simulation run number in figure~\ref{fig5ipvsn}(a) for both values of $Z_{eff}$. Current in the first run at which  only the fundamental harmonic is involved comes out zero due to no resonance occuring between the wave vector and electron velocity around the $n=1$ harmonic layer. Being the dominant one in terms of power deposition, 2nd harmonic in the 2nd run amounts to 6.2 kA current for $Z_{eff}=2.5$ and 6 kA for $Z_{eff}=4.0$. As we add on higher harmonics, $I_p$ for $Z_{eff}=2.5$ (green line) and $Z_{eff}=4.0$ (red line) quickly ramps up with contribution from harmonics $n=3-5$. Power and current density
\begin{figure}[htbp]
\subfloat[]{\includegraphics[height=6.0cm,width=8.5cm]{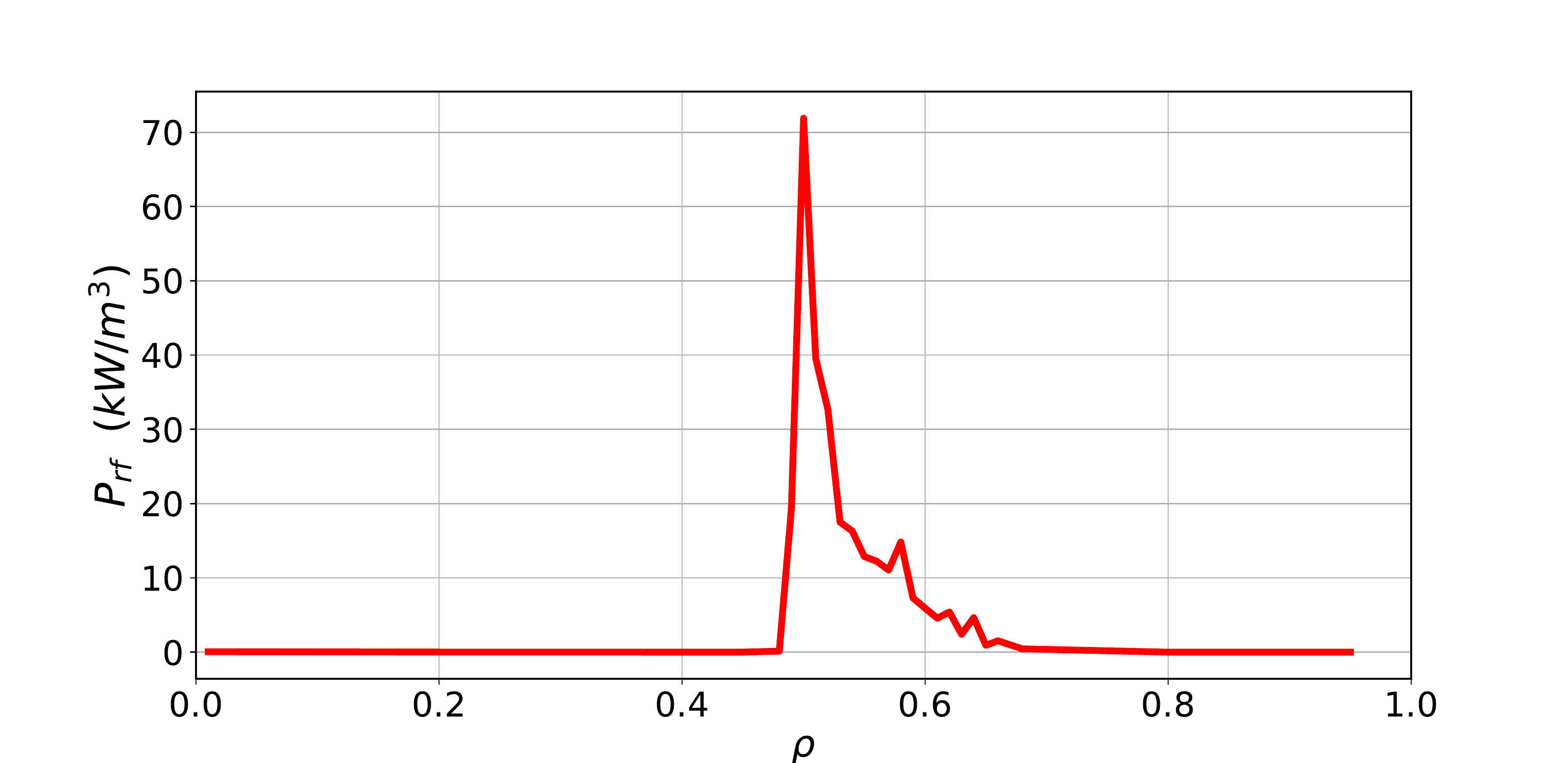}}
\subfloat[]{\includegraphics[height=6.0cm,width=8.5cm]{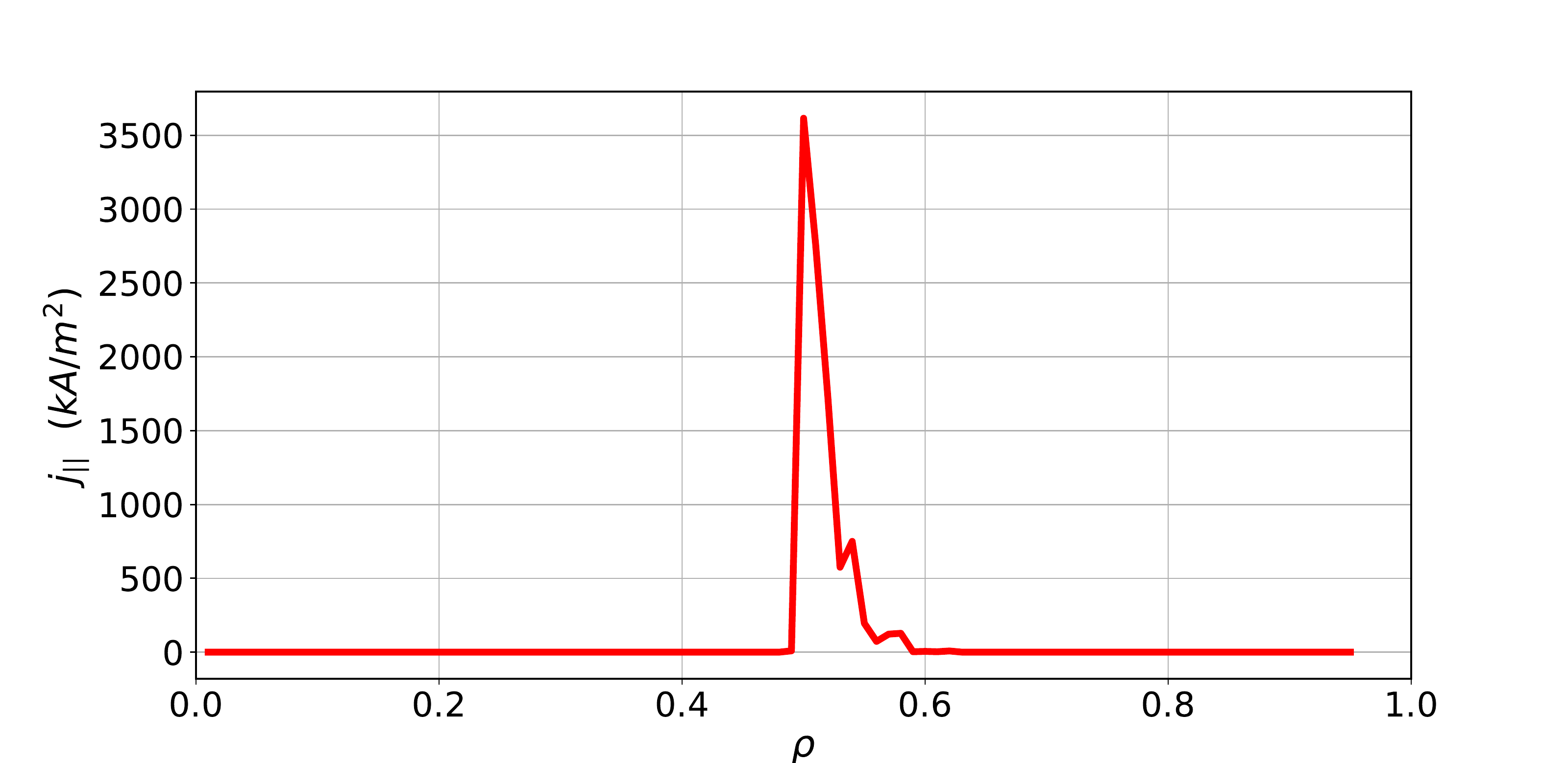}}%
	\caption{\label{fig6prf} Power deposition (a) and current density (b) with minor radius in case of X-wave from the run $n=1-5,~ Z_{\rm eff}=4.0$. Both profiles are localized around the cold 2nd harmonic ECR layer.   }
\end{figure}
profiles are seen as localized around the 2nd harmonic cold ECR location within $\rho=0.48-0.6$, even for the cases those include harmonics higher than $n=2$ in calculation (figure~\ref{fig6prf}). It may be interpreted that, as per the ECRH theory, electrons moving around the cold 2nd ECR location resonate with it to absorb high amount of wave power. Associated with this, a high value of the asymmetric QL diffusion generates long tail of electron velocity distribution via the pitch angle scattering of long range Coulomb interaction. Thus created energetic electrons eventually resonate with higher harmonics ($n>2$) by means of a two-step mechanism; first, through the relativistic Doppler shifted resonance and second, via a nonlinear interaction with different harmonics. The first step is widely interpreted as being effective for power absorption in fully EC wave powered operation on QUEST~\cite{onchi21}. This article explores the nonlinear second step that comes owing to the quadratic dependence of QL diffusion coefficients on wave electric field. The sharing of total absorbed power  among different harmonics in every run is displayed in figure \ref{fig5ipvsn}(b). A large fraction of power goes to the 2nd harmonic in each run with a gradual drop in its value as more number of harmonics are getting involved. 
\begin{figure}[htbp]
\subfloat[]{\includegraphics[height=6.0cm,width=8.5cm]{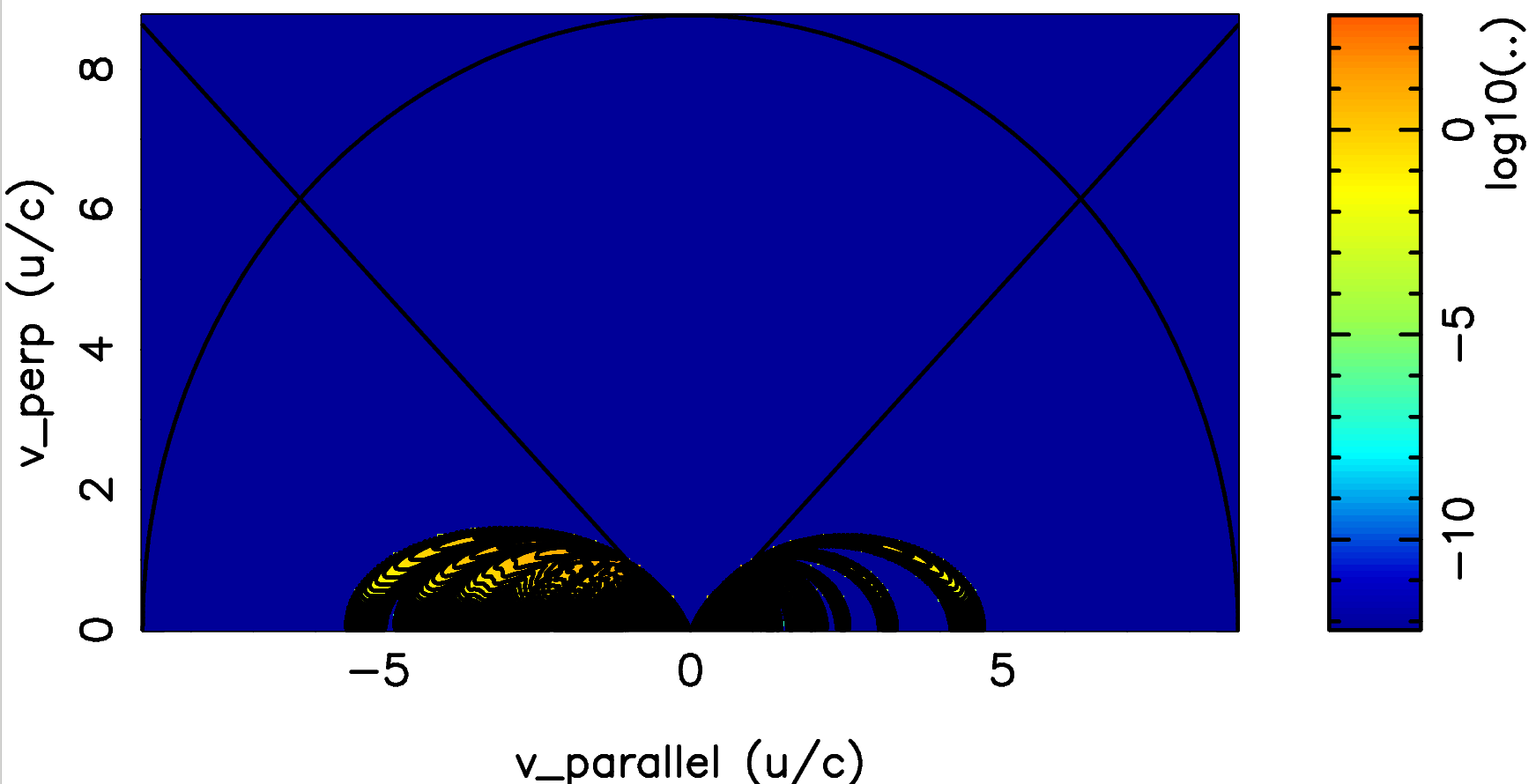}}
\subfloat[]{\includegraphics[height=6.0cm,width=8.5cm]{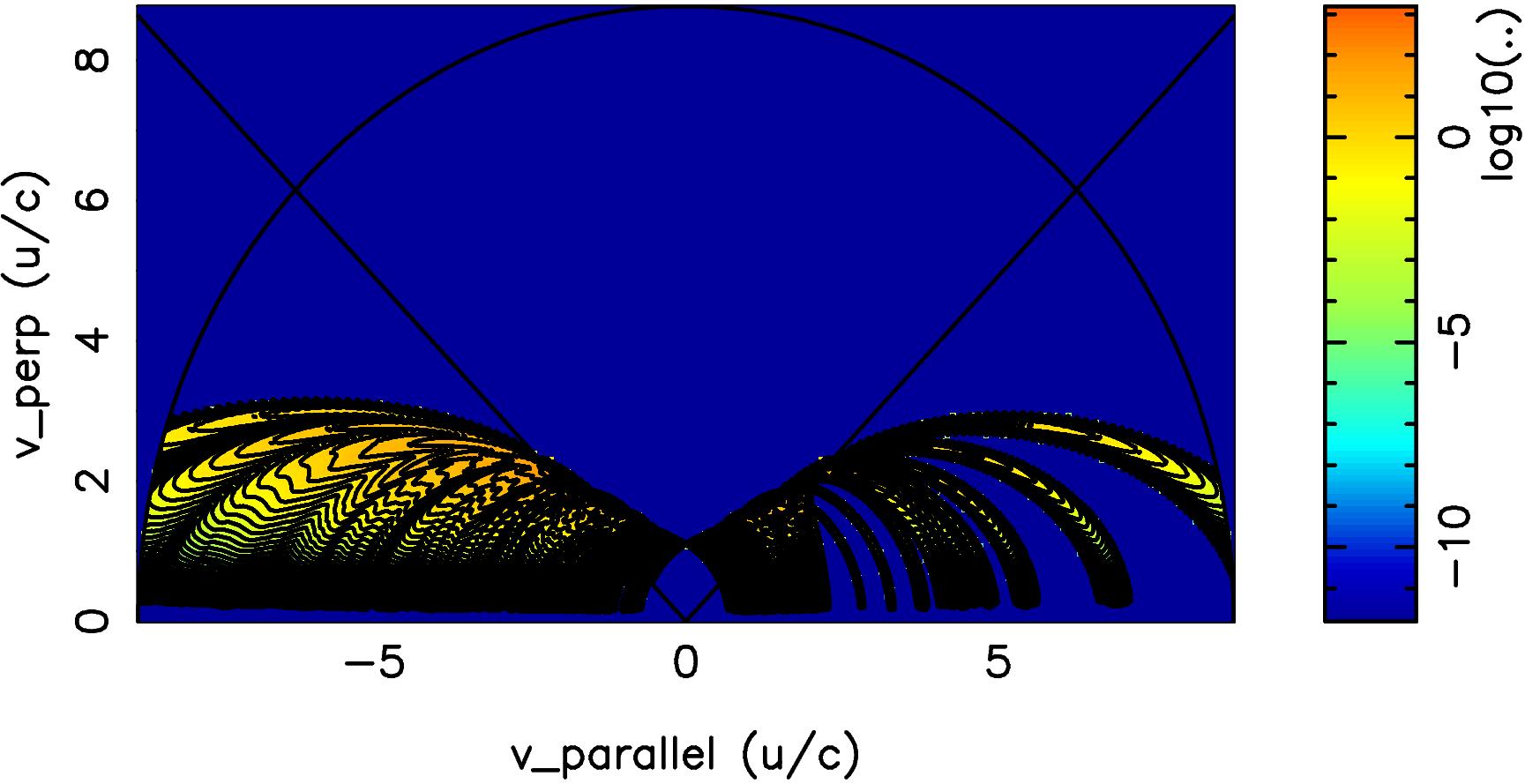}}
	\caption{\label{fig7diffx} (a) Quasi-linear diffusion coefficient in relativistic ${\bf u}/c \equiv {\bf p}/mc$ space computed with CQL3D in case of X-wave for the 3rd harmonic (a) and for the 5th harmonic (b) from a run $n=1-5, ~ Z_{\rm eff}=4.0$. The diffusion is generally high away from the t-p boundary (solid black straight lines), but $n=5$ harmonic exhibits diffusion across the t-p boundary into trapped particle region.  }
\end{figure}

\begin{figure}[htbp]
\subfloat[]{\includegraphics[height=6.0cm,width=9.0cm]{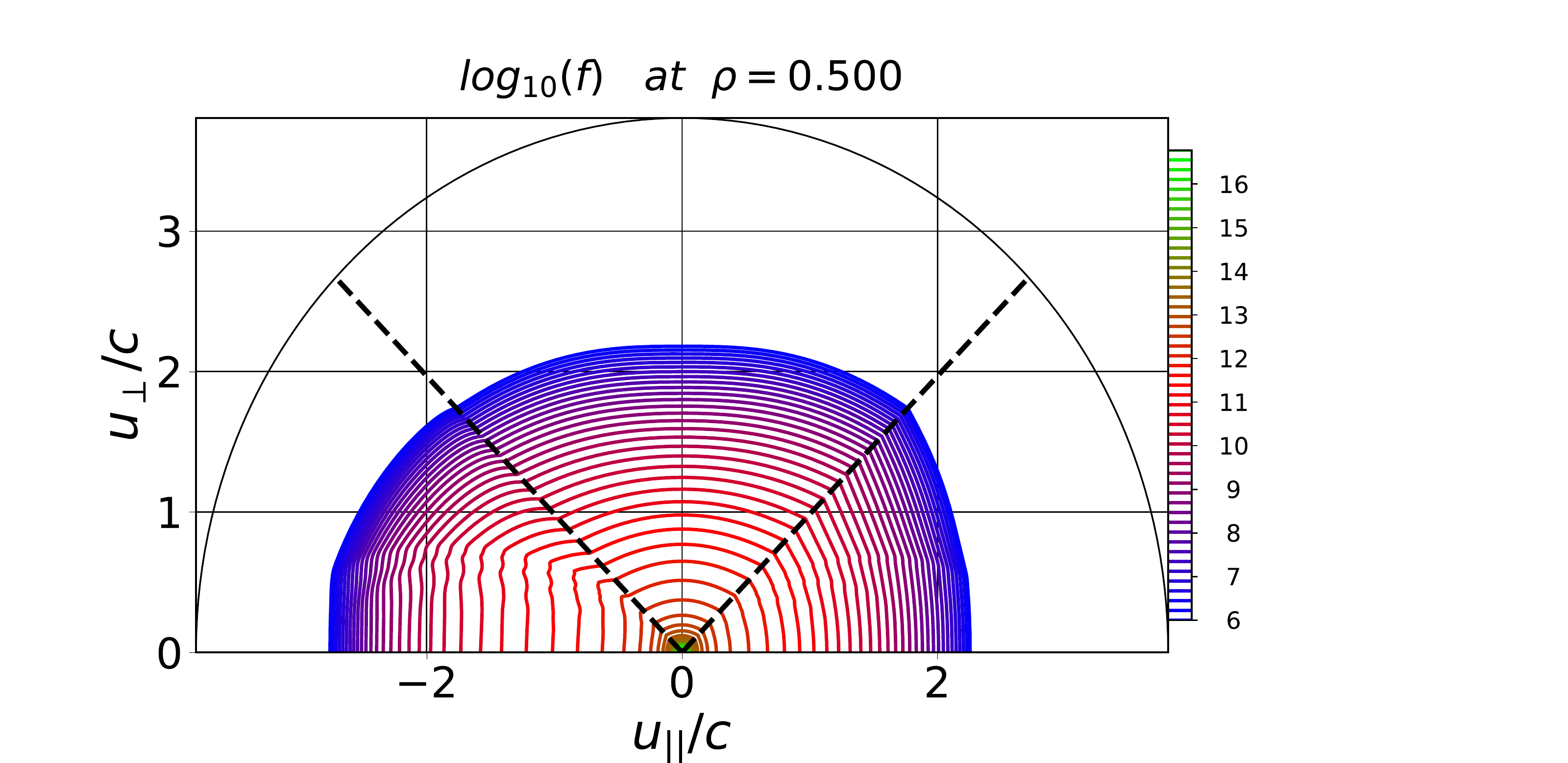}} 
\subfloat[]{\includegraphics[height=6.0cm,width=9.0cm]{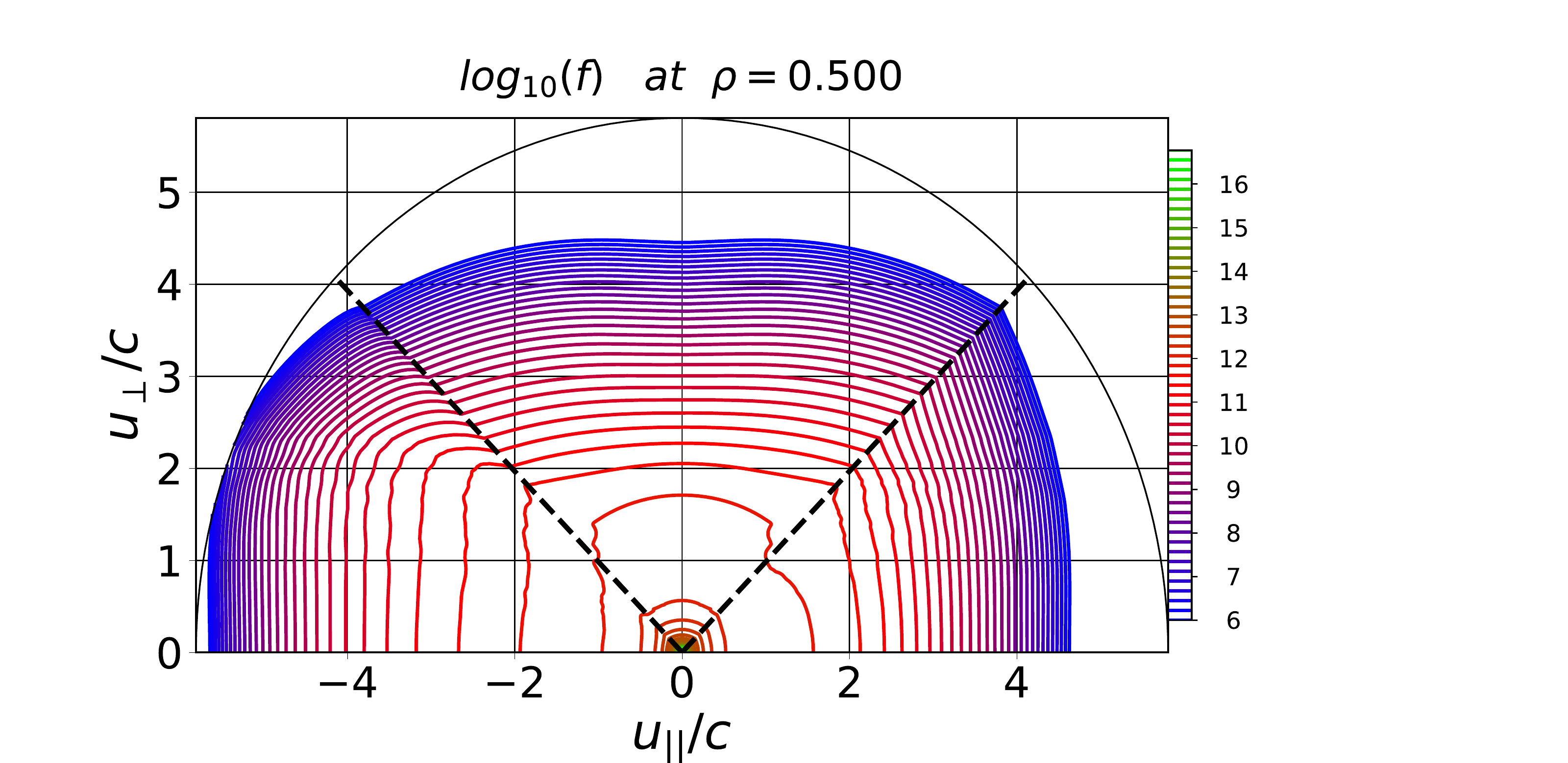}} 
	\caption{\label{fig8distxs} Electron distribution function ($f_e$) in relativistic ${\bf u}/c \equiv {\bf p}/mc$ space computed with CQL3D for the X-wave; (a) $f_e,~ n=1-2,~ Z_{eff}=4.0$;~ (b) $f_e,~ n=1-5,~ Z_{eff}=4.0$. Both distribution functions are fetched from minor radius location $\rho=0.5$. }
\end{figure}
\begin{figure}[htbp]
\subfloat[]{\includegraphics[height=5.6cm,width=5.5cm]{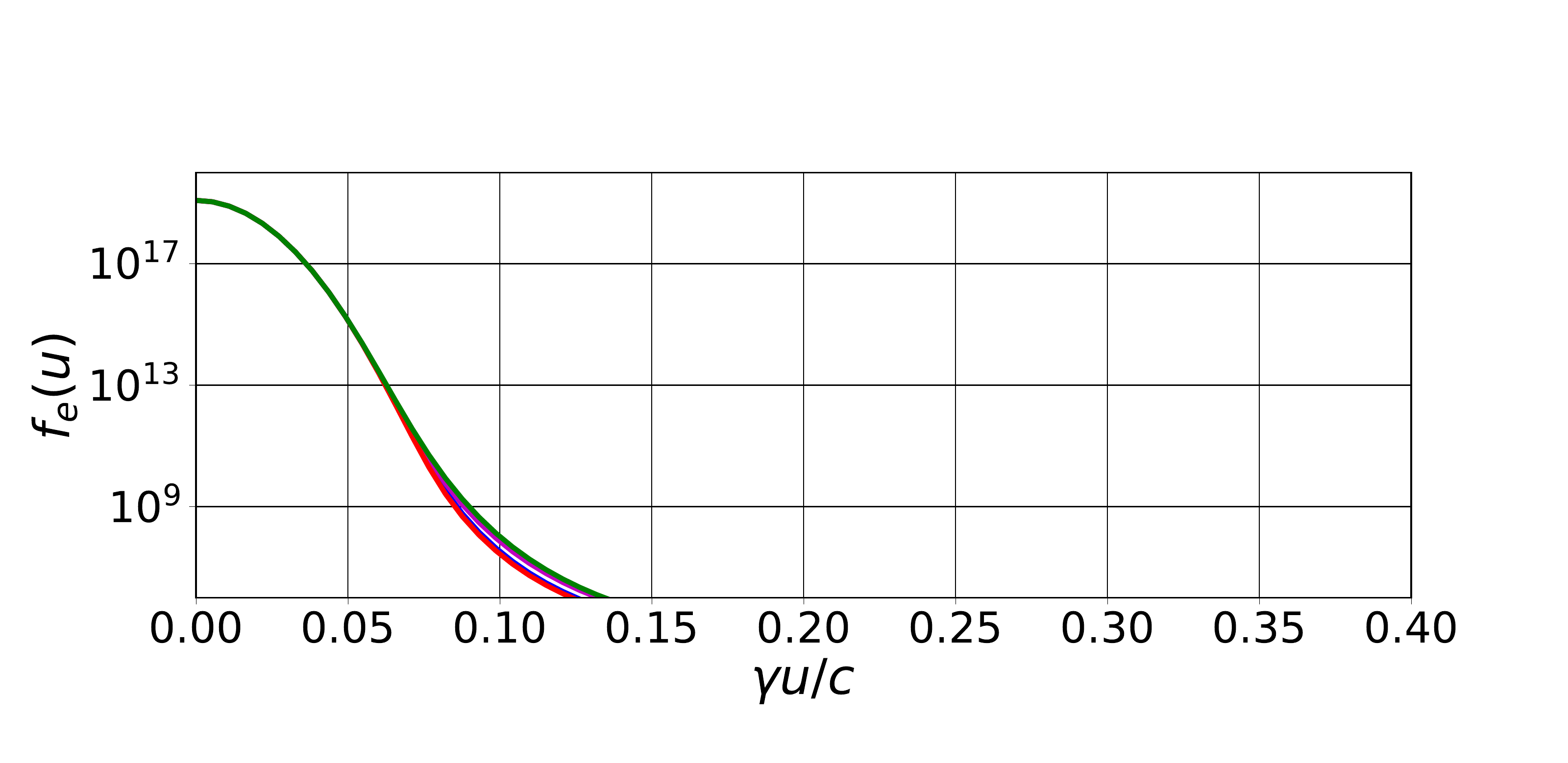}}
\subfloat[]{\includegraphics[height=5.2cm,width=5.5cm]{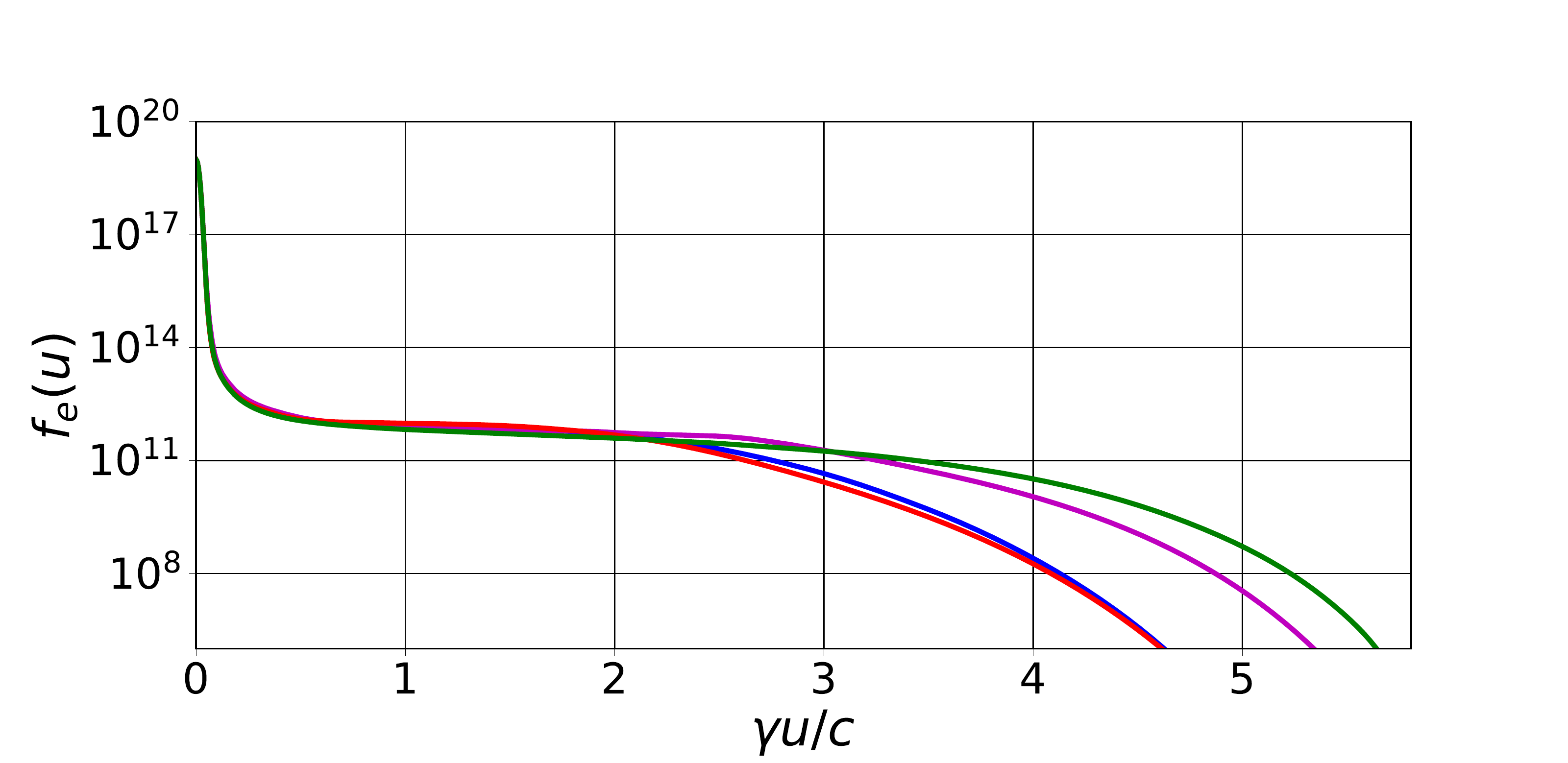}}
\subfloat[]{\includegraphics[height=5.6cm,width=5.5cm]{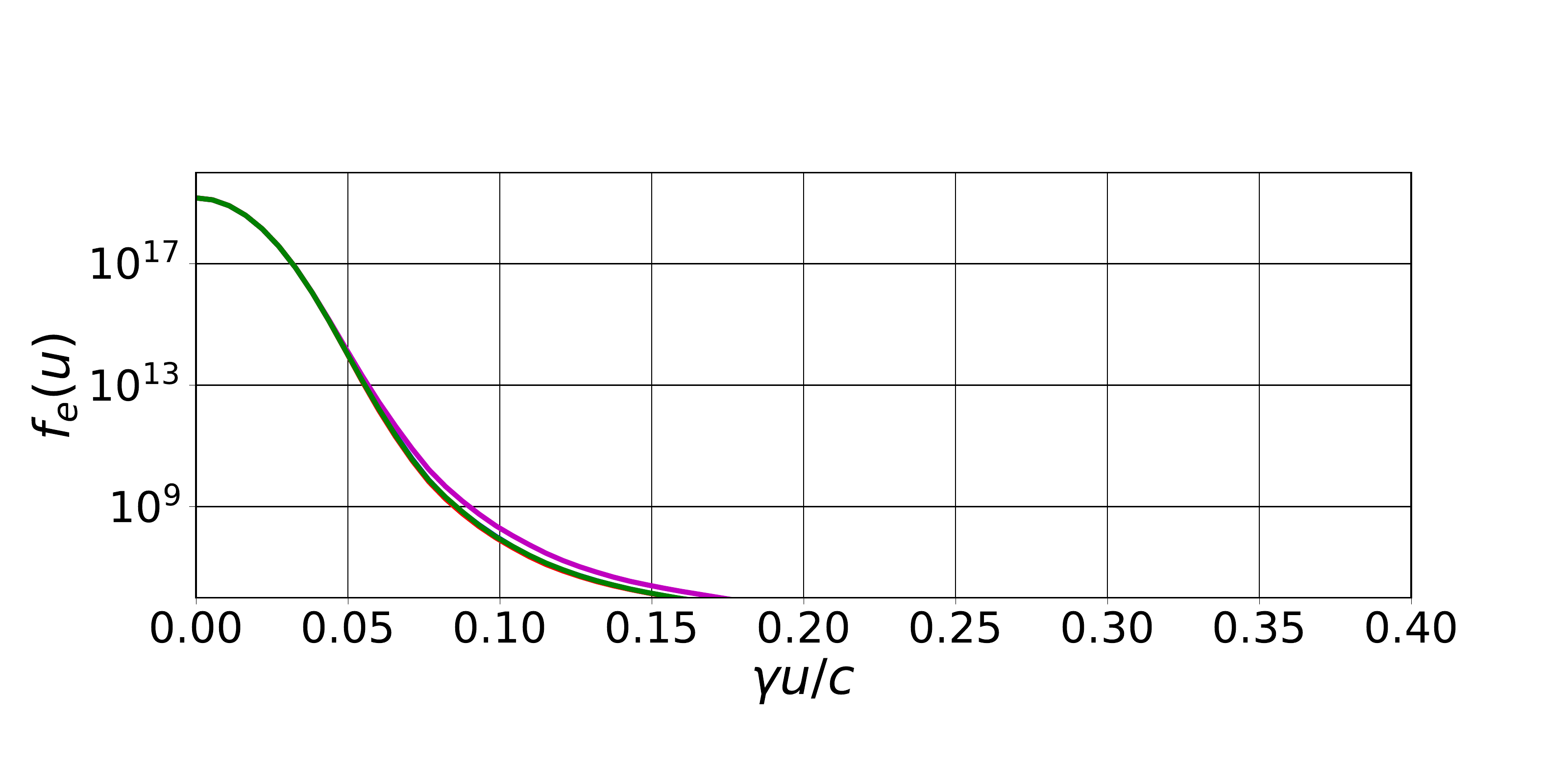}}\\
\caption{\label{fig9fju}cuts at four different constant pitch angles through the 2D electron distribution function $f_e$ taken from three different radial locations (a) $\rho=0.01$, (b) $\rho=0.5$, (c) $\rho=0.9$ vs. $\gamma u /c$ , where $\gamma$ is the relativistic factor. $\rho=0.5$ position characterizes a long tail of distribution compared to other two locations.}
\end{figure}

In CQL3D, $Z_{eff}$ appears in the Coulomb collision calculation between electrons and different impurity ion species. The QL diffusion coefficient in 2D velocity space is plotted in figure \ref{fig7diffx} for the harmonics third and fifth from a simulation run for $n=1-5$ harmonics. Figure \ref{fig7diffx}(a) clearly shows diffusion is more evident in passing particle region away from the trapped-passing (t-p) boundary, and it is asymmetric being dominant for velocity opposite to local magnetic field. 
Diffusion property changes noticeably for the 5th harmonic as it enters trapped region by crossing t-p boundary, and further extends to high velocity regime (figure \ref{fig7diffx}(b)). Important characteristic to note is the maximum diffusion location (intense coloured region in plots) moves further to  higher velocity region with the increasing order of ECR harmonic. In figure \ref{fig8distxs}, distribution functions in velocity space are shown for the two runs $n=1-2$ and $n=1-5$. There is clear asymmetry in high velocity region and the most distinguishable asymmetric line in distribution for the $n=1-5$ case is lying more close to the t-p boundary compared to the $n=1-2$ case. 
From a case study of $Z_{eff}=4.0$ and $n=1-5$, we have demonstrated the different nature of distribution functions at three radial locations - at $\rho=0.01$ near to the 3rd harmonic cold ECR (figure~\ref{fig9fju}(a)), at $\rho=0.5$ near to the 2nd harmonic cold ECR (figure~\ref{fig9fju}(b)), at $\rho=0.9$ near to the 4th harmonic position (figure~\ref{fig9fju}(c)). Clearly, $\rho=0.5$ location contains a long tail of high velocity electron population in addition to a low temperature bulk thermal population. But, no such tail is visible in electron distribution at $\rho=0.01$ or $\rho=0.9$. It is due to weak resonance between thermal electrons and the 3rd \& 4th harmonic ECRs at low temperature ($<100$ eV).
%
\newpage
\subsection{X-wave simulation with input power scan}
\begin{figure}[htbp]
\subfloat[]{\includegraphics[height=7.0cm,width=12.0cm]{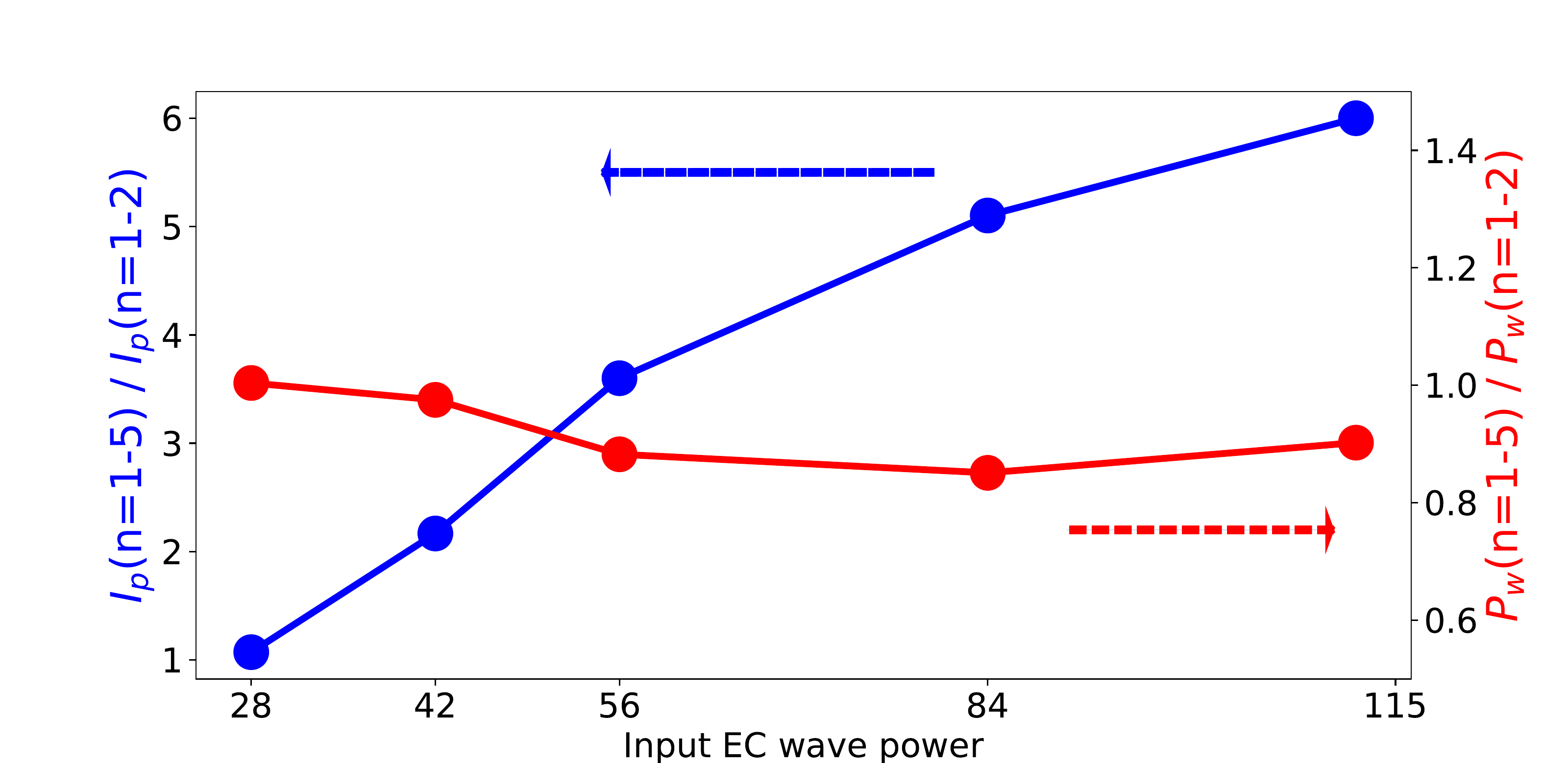}}
	\caption{\label{fig10non}Ratio of absorbed power ($P_w(n=1-5)/P_w(n=1-2)$, red line) and current ($I_p(n=1-5)/I_p(n=1-2)$, blue line) from the $n=1-5$ run to that of $n=1-2$ run. With rise in input power, strong nonlinearity increases the ratio of $I_p$, but makes power ratio dipped slightly. }
\end{figure}
In order to demonstrate the non-linear nature of this property of multi-harmonic influence on current drive, we scan the injected ray power so that the QL diffusion coefficient which is quadratic in wave electric field changes. Input EC power in CQL3D is step-wise dropped from 115 kW down to 28 kW, and simulation is carried out for every power value to obtain absorbed power ($P_W$) and $I_p$ for two cases; one run with two harmonics $n=1-2$, other run with 5 harmonics $n=1-5$. Then, the ratios are made for CQL3D output $P_W$ and $I_p$ values between the $n=1-5$ and $n=1-2$ runs, and plotted in figure \ref{fig10non}. This scan doesn't alter power deposition much but it has a profound effect on current drive. The blue line labeled with the ratio of $I_p$ from the $n=1-5$ run to that from the $n=1-2$ run infers that, at low 28 kW input RF power, higher harmonics have no noticeable effect, but with rise in input EC wave power the ratio increases non linearly as higher harmonics contribute. A slight reduction noticeable in power ratio (the red line curve) with increase of input EC wave power may be understood as follows; with higher power applied, more fraction of electrons resonate with higher harmonics ($n>2$), and consequently total power absorption drops as we know ECR damping on higher harmonics is weaker than on the second.   
%
%
\subsection{X-wave simulation with density and temperature scan}
\begin{figure}[htbp]
\includegraphics[height=8.5cm,width=12.5cm]{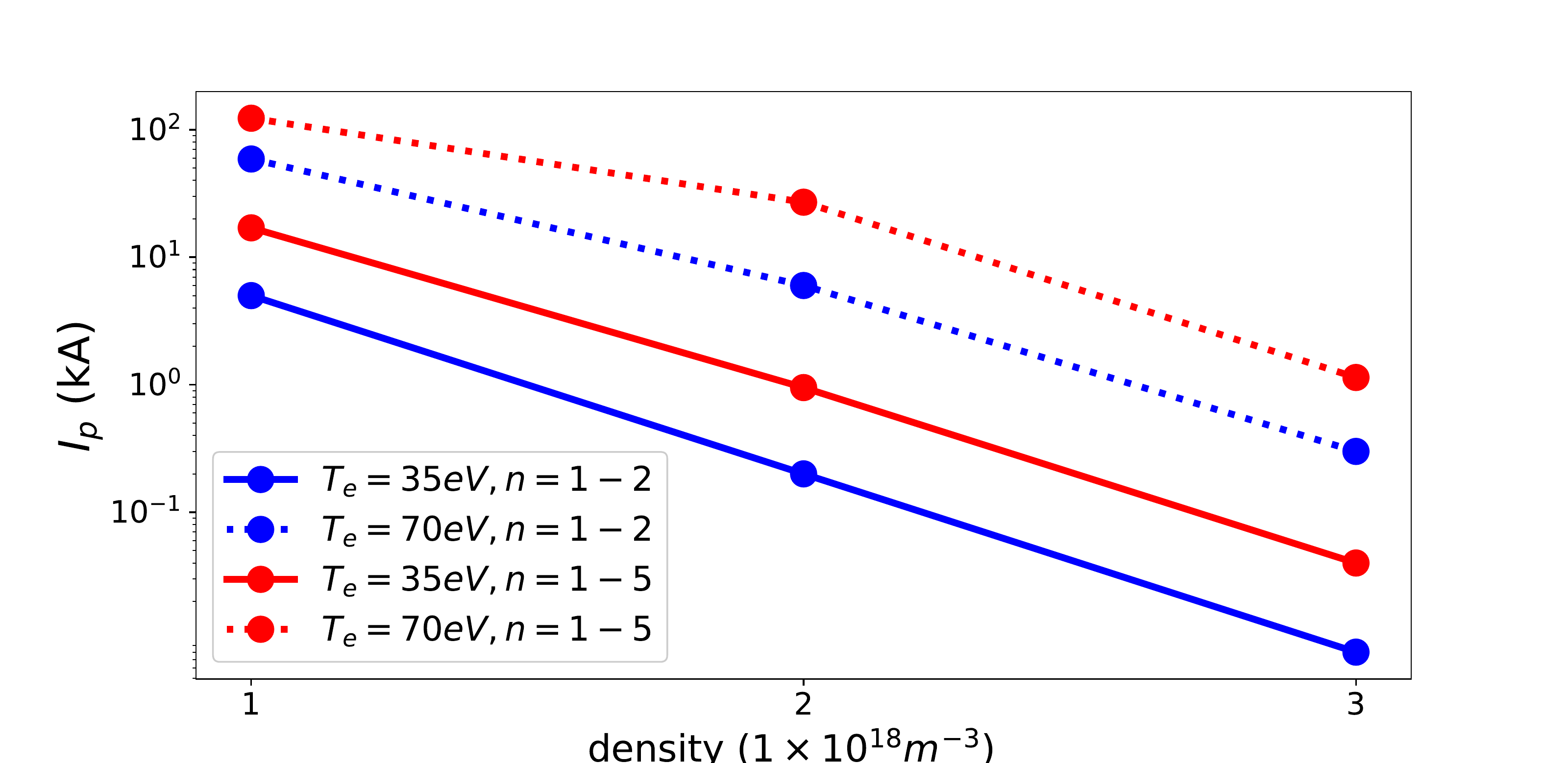}	
\caption{\label{fig11dt} $I_p$ vs density for two values of $T_{e}$ at the cold 2nd ECR location from the two runs $n=1-2$ and $n=1-5$. Results of $T_e = 35$ eV run are drawn as solid lines, and that of $T_e=70$ eV by dotted lines. In all scenarios, the $n=1-5$ run exhibits higher $I_p$ compared to the $n=1-2$ run. }
\end{figure}

Density and temperature at the cold 2nd ECR location ($\rho=0.5$) is crucial as the process of wave power absorption and correspondingly energetic electron generation is sensitive to these parameters. It is important to know  whether the effectiveness of multi-harmonic ECR sustains in different values of these two parameters. To address this, we have decided to continue simulation with a lower and a higher values of plasma density, and two different temperature values at $\rho=0.5$, and compared the results. In figure \ref{fig11dt}, drawn in logarithmic ordinate and normal abscissa, $T_{e}=35$ at $\rho=0.5$ case results are denoted by solid line curves, and $T_{e}=70$ at $\rho=0.5$ by dotted line curves for two runs - one run with $n=1-2$ harmonics and other with $n=1-5$ harmonics. We have found that the utility of multi-harmonic ECRs, quantified as the ratio of $I_p$ obtained in simulation run with $n=1-5$ harmonics over that of $n=1-2$ harmonics, reduces by half when density is lowered from $n_{e}=2 \times 10^{18}~m^{-3}$ to $n_{e}=1 \times 10^{18}~m^{-3}$ for both temperature case studies. However, for any particular density, this effectiveness hasn't changed much with temperature. It is possibly because the energetic electron population resonating with $n>2$ harmonics reduces at lower $n_e$. Obviously, $I_p$ drops if $T_{e}$ is lowered due to the reduction in power absorption by electrons at lower temperature. Still a low parameter scenario like as $n_{e}=1\times10^{18}~m^{-3}$ and  $T_{e}=35$ eV simulates about 17 kA current on basis of single-pass power absorption, a sufficiently effective value. At higher density, current drops but the effectiveness of multi-harmonic ECRs persists. The exponential fall of $I_p$ with $n_e$ is noticeable, and this is further discussed in section~5 in connection with high density experiments on EXL-50. 
\newpage
\section{Modeling of Multi-pass absorptions with the 28 GHz EC wave reflecting on wall}
\begin{figure}[htbp]
\includegraphics[height=8.5cm,width=6.5cm]{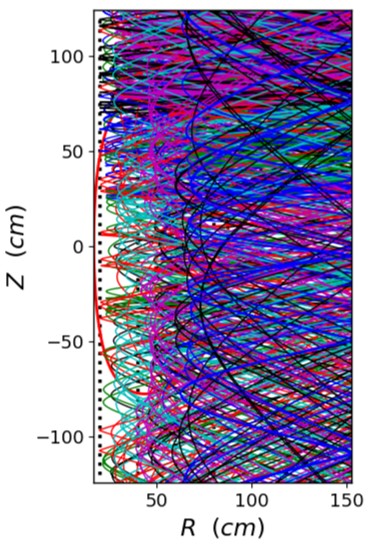}	
\caption{\label{fig12mrays} Ray-tracing diagram in poloidal plane from the multi-pass X-mode simulation with 36 times reflection on wall. Rays have covered entire 2D space including both open and closed field line regions.}
\end{figure}
As explained in section~(2), a multi-pass absorption simulation is inevitable to fully
represent a low density, low temperature experimental scenario of EXL-50. The GENRAY code's capability of tracing multi-pass ray propagation with ray-reflection occurred on simulation boundary (representing vessel wall) is implemented in this section. Now, rays are visible to be in both inside and outside  the LCFS
 basically filling up nearly the full volume of simulation domain as drawn in figure \ref{fig12mrays}. 
\begin{figure}[htbp]
\subfloat[]{\includegraphics[height=7.0cm,width=9.0cm]{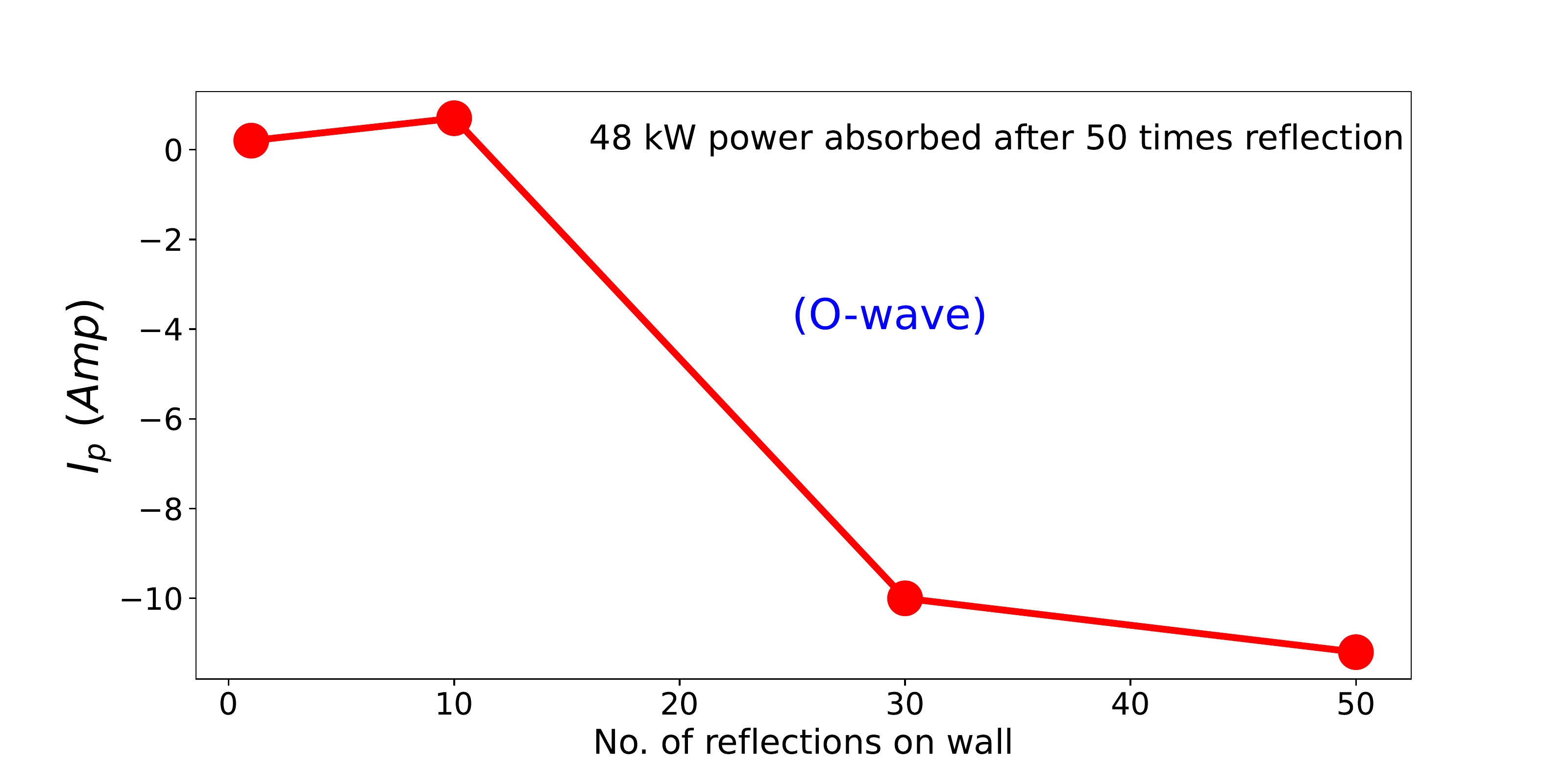}}
\subfloat[]{\includegraphics[height=7.0cm,width=9.0cm]{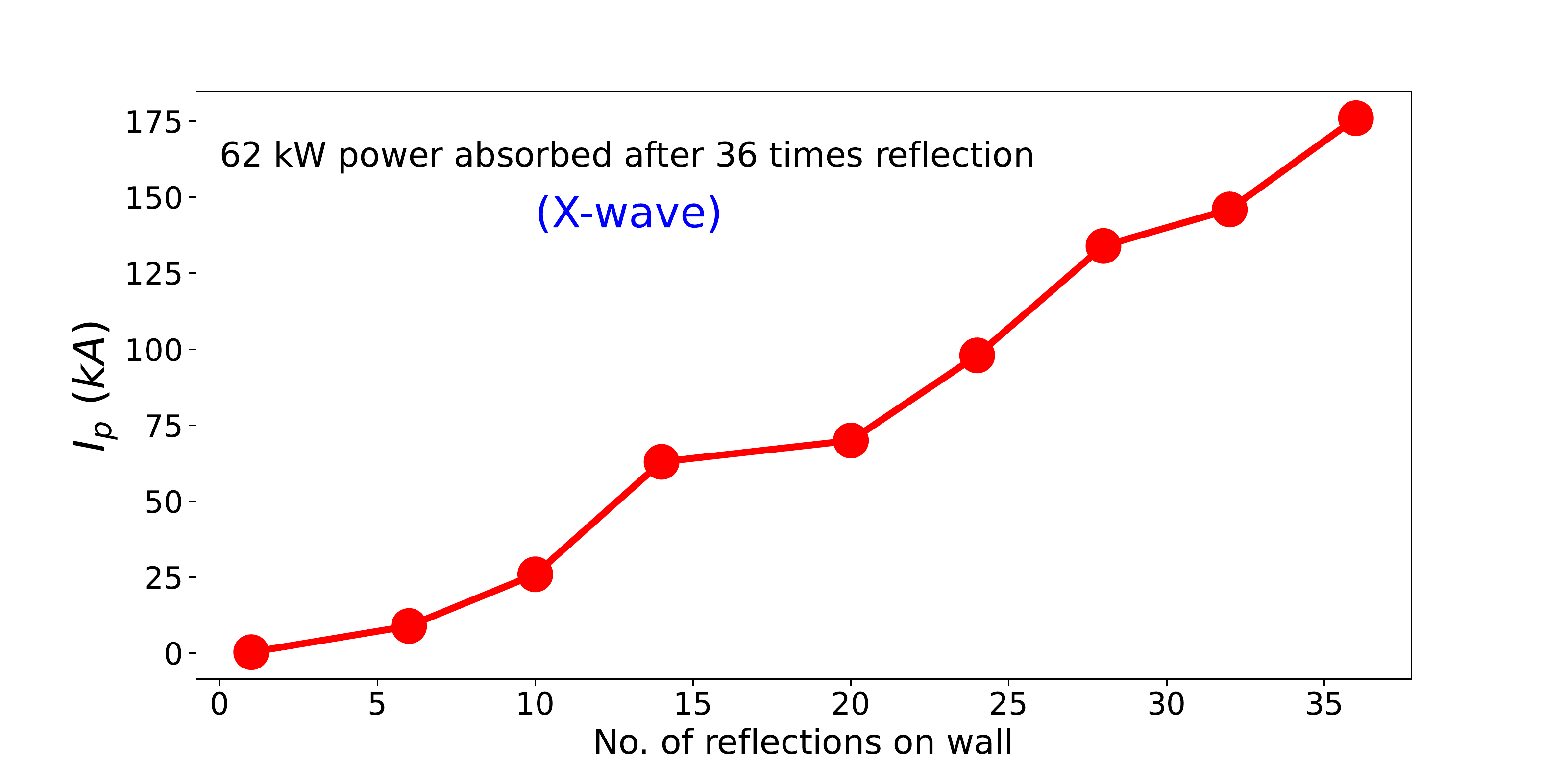}}%
\caption{\label{fig13multiip} (a) $I_p$ vs. number of reflection on wall for O-wave multi-pass runs. The generated current in this case is of negligible amount eventhough in total $\sim 48$ kW power has been absorbed. (b) $I_p$ vs. number of reflection on wall for X-mode multi-pass runs. $I_p$ rises with step wise increase of reflection numbers finally reaching to $\sim 175$ kA level in expense of $\sim 62$ kW input power.   }
\end{figure}
However, CQL3D being a bounce averaged code in poloidal direction can only deal with a closed field line plasma. Therefore, current flowing along open field lines in EXL-50 can't be simulated with it at present. Nevertheless, this section may help us understand current drive mechanism inside the closed field plasma region for multi-pass absorptions. Like single-pass ray-tracing in section (3), magnetic field profiles are fetched from discharge $\#7672$, density and temperature profiles are set like as figure \ref{fig2eqx} with only change in the core density value $n_{eo}=1\times10^{18}~m^{-3}$, aiming for the low line integrated density value of discharge $\#7672$. 

Multi-pass simulations are conducted separately for both O- \& X-waves with injecting 115 kW input power from the ECRH2 launcher for each case. Since the wave polarisation may change from being O- to X- mode type or vice versa upon reflection on the device wall, and such mode conversion is not currently modeled in GENRAY, two sets of multi-pass simulation are carried out separately for the O- \& X-wave by gradually increasing number of reflections until nearly half of the input power is found absorbed for each wave. 
Figure \ref{fig13multiip}(b) provides us the information that about 175 kA current is driven in expense of 62 kW input power absorbed, after as many as 36 times reflection on simulation boundary have occurred for the X-mode EC wave.
\begin{figure}[htbp]
\subfloat[]{\includegraphics[height=6.0cm,width=9.0cm]{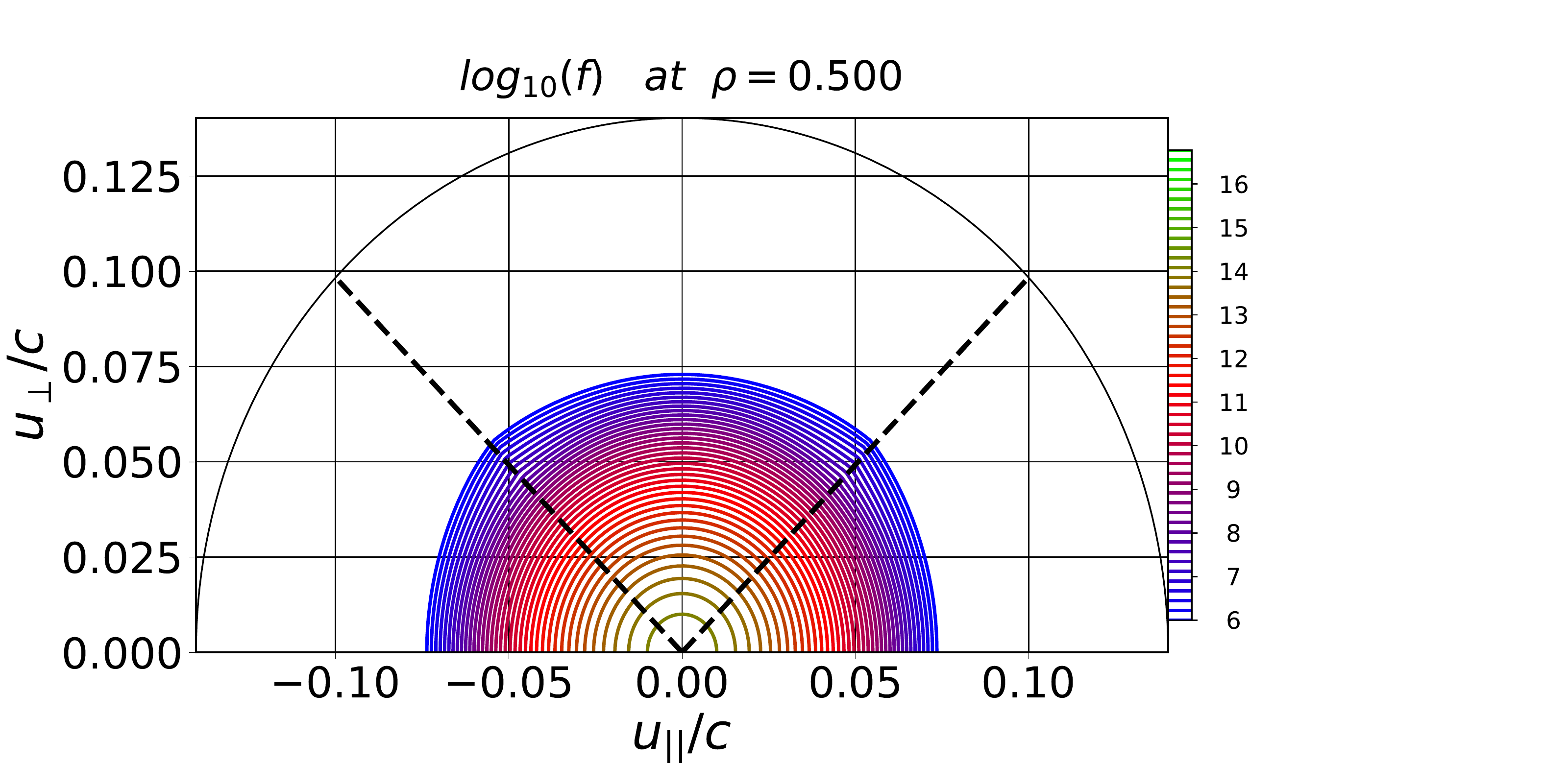}}
\subfloat[]{\includegraphics[height=6.0cm,width=9.0cm]{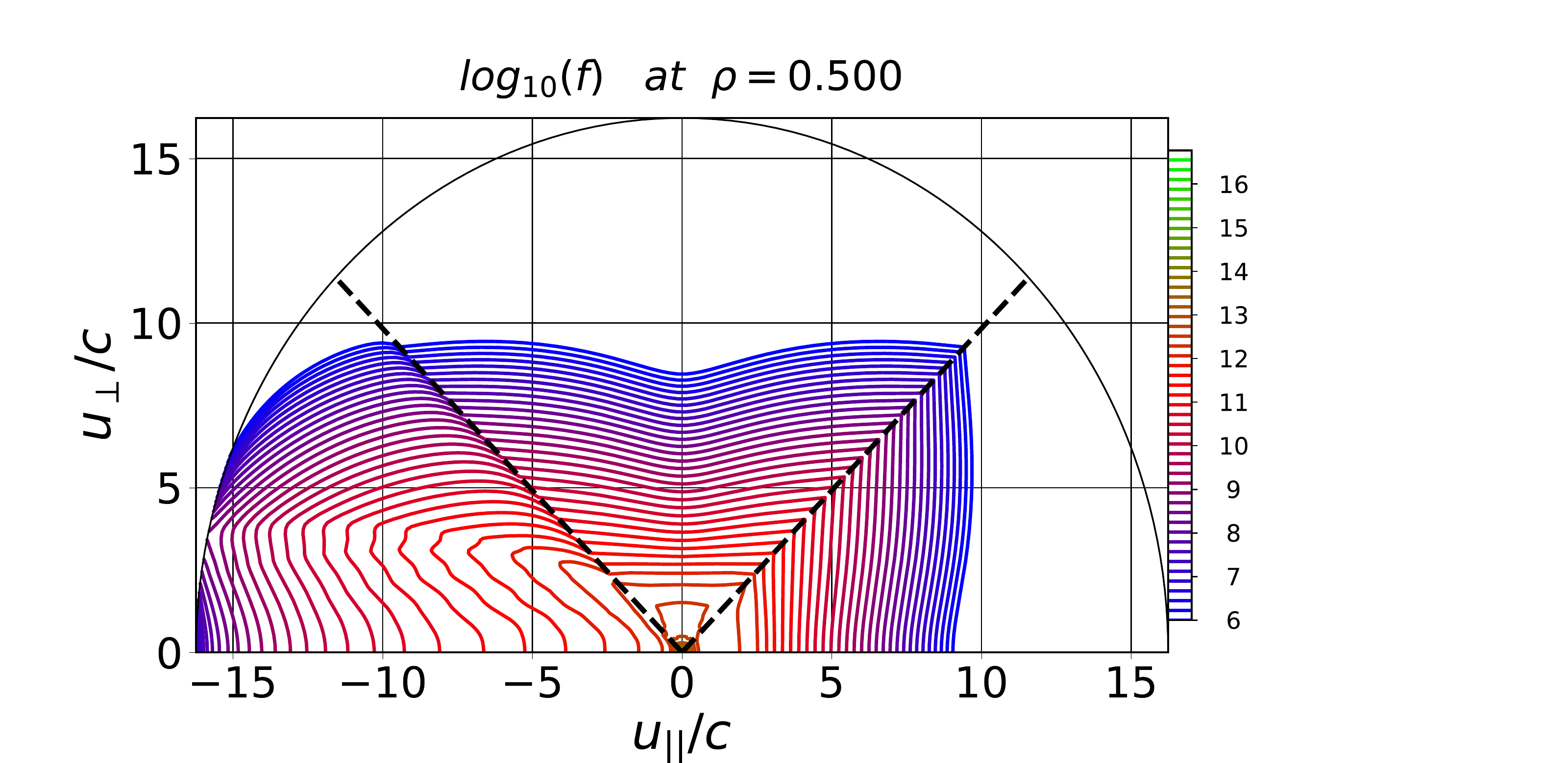}}\\
\caption{\label{fig14multi} Electron distribution function ($f_e$) in relativistic ${\bf u}/c \equiv {\bf p}/mc$ space computed with CQL3D for multi-pass absorption. (a) $f_e$ from the O-mode run performing 50 times reflection, (b) $f_e$ from the X-mode run performing 36 times reflection on the simulation boundary.}
\end{figure}
On other hand, a merely 10 Amp counter current is flown in the case of O-wave while 48 kW power is absorbed after 50 times reflection.  Here, total absorbed power includes power due to collisional damping and reflection loss on wall in GENRAY alongwith the power deposition via wave-electron resonance. Even though this cumulative current is a little exaggeration of the experimental steady state current 125 kA achieved in shot $\#7672$, the order of estimation might be acceptable given the fact that real power sharing between the O-wave and X-wave is hard to speculate in multi-pass scenario in experiment, and the precise values of density, temperature and $Z_{\rm eff}$ at resonance locations are not available at present. From the contour plots of electron distribution function at the final stage of simulation for both O- \& X-waves in figure \ref{fig14multi}(a,b), we may understand the reason that makes these two cases so different from each other, as the X-wave diffusion is highly asymmetric whereas the O-wave diffusion is almost symmetric in co- and counter-direction of $u_{\parallel}/c$. 
\section{Discussions}
\begin{figure}[htbp]
\subfloat[]{\includegraphics[height=6.0cm,width=8.0cm]{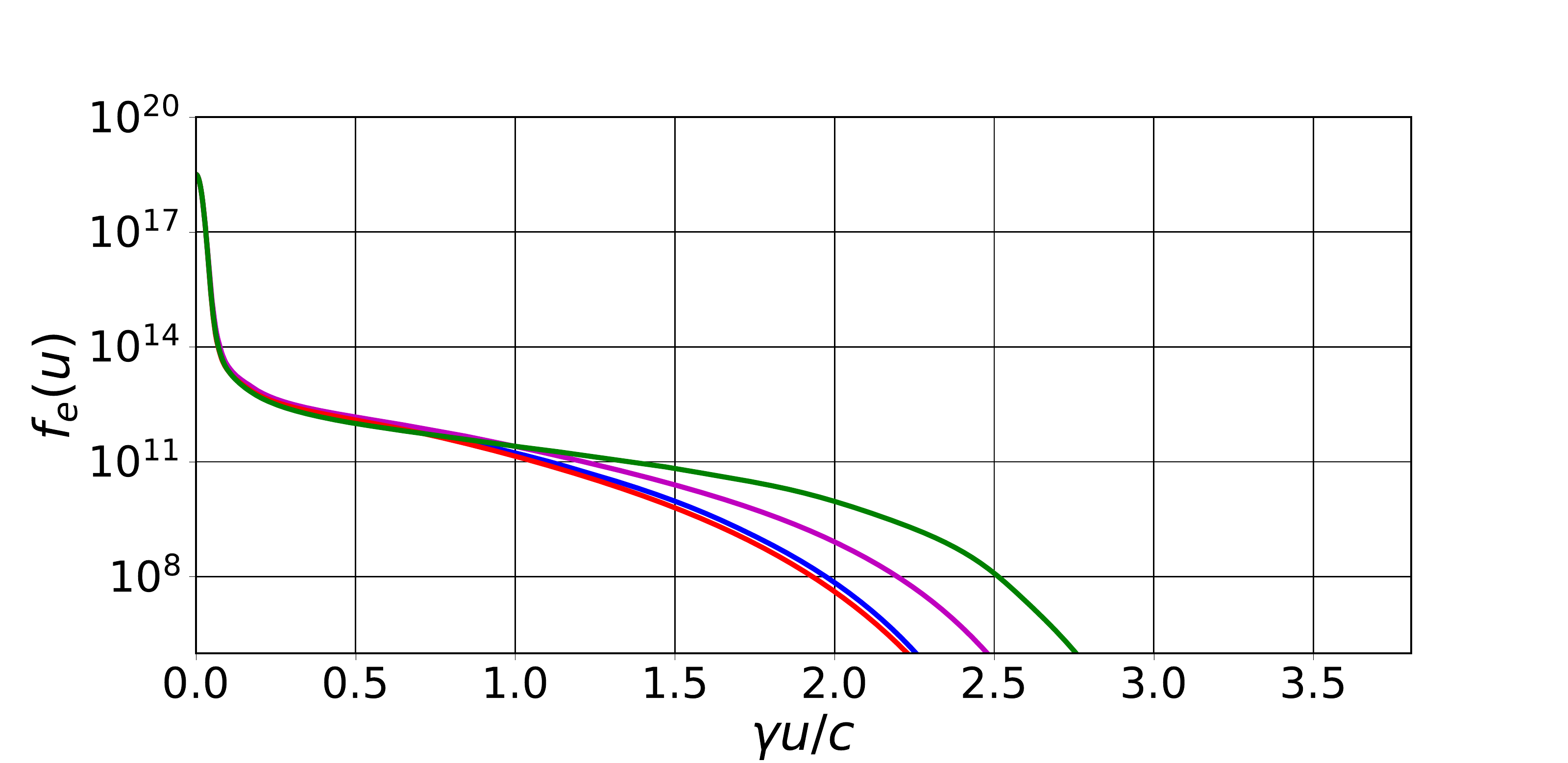}}
~~\subfloat[]{\includegraphics[height=6.0cm,width=8.0cm]{dist1d5.eps}}\\
\subfloat[]{\includegraphics[height=6.3cm,width=8.0cm]{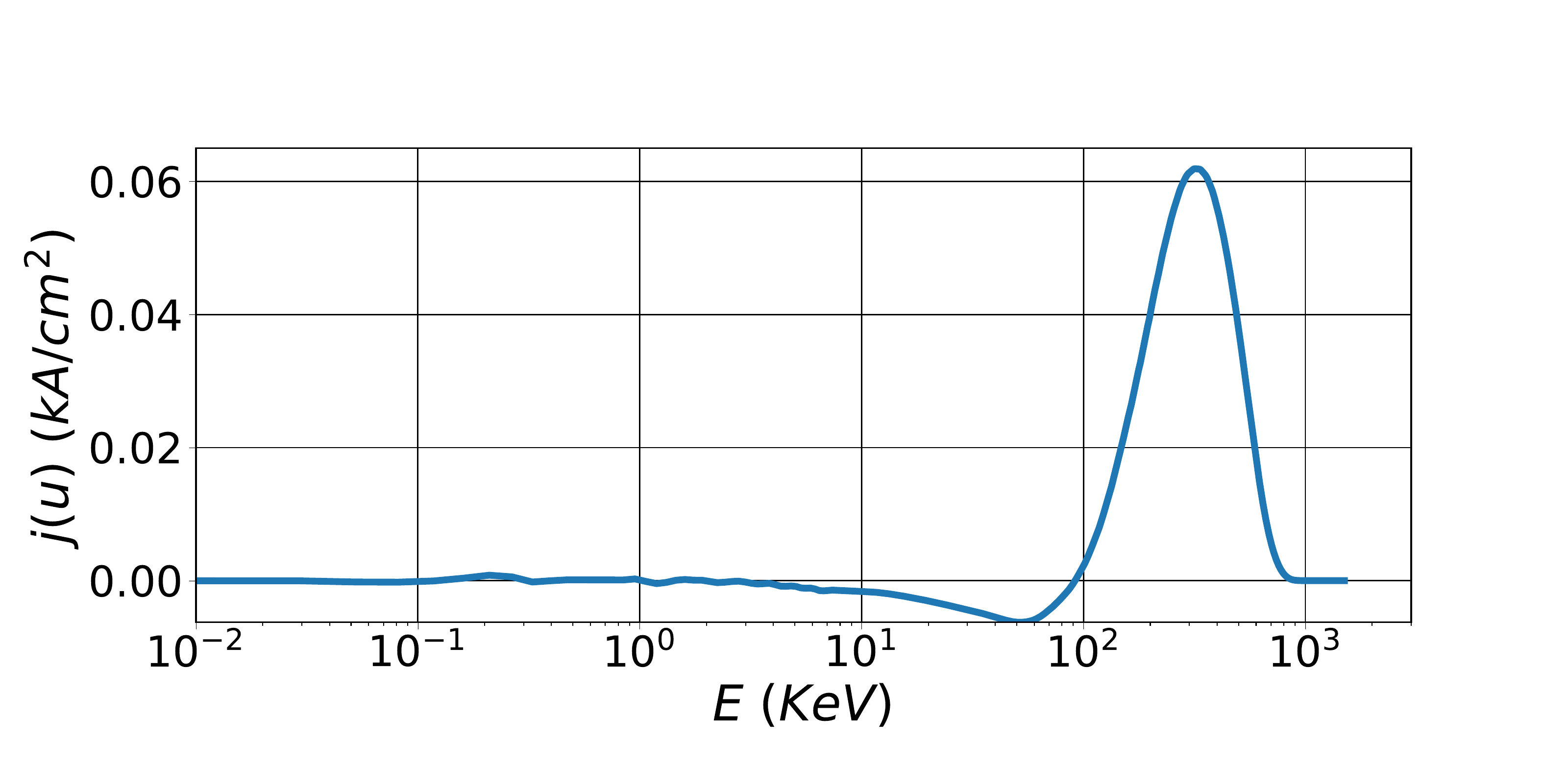}}
~~\subfloat[]{\includegraphics[height=6.0cm,width=8.0cm]{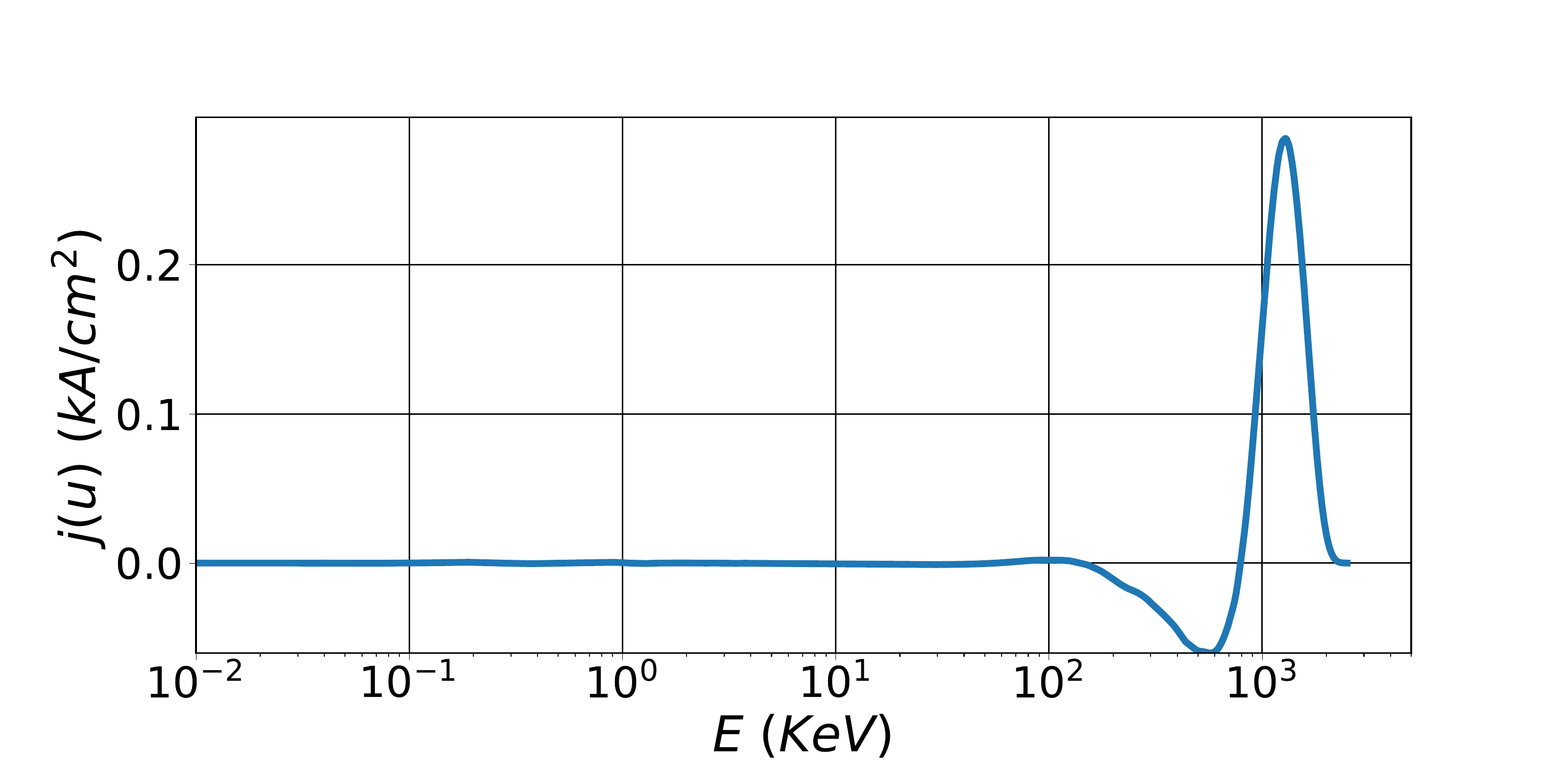}}%
	\caption{\label{fig15fju} Cuts at four different constant pitch angles through the 2D electron distribution function $f_e$ taken from two different case studies - (a) $n=1-2,~ Z_{\rm eff}=4.0$ (b) $n=1-5,~ Z_{\rm eff}=4.0$ vs. $\gamma u /c$ , where $\gamma$ is the relativistic factor. For same runs current density function $j(u)$ vs. energy of electrons for $n=1-2$(c) \& $n=1-5$(d). Total current density at respective radial position is $\int j(u) du$ with $u$ being in normalized unit.}
\end{figure}
In the X-wave single-pass simulation, inclusion of multiple harmonics in calculation  is found to strengthen current drive efficiency of the EC wave. This has happened because high velocity electrons around the cold 2nd ECR location ($\rho=0.5$) resonate strongly with higher harmonics ($n>2$) as per the relativistic Doppler shifted resonance condition to become more  and more energetic and, as a result, drive higher current. We can visualize this fact in figure \ref{fig15fju}(a,b) where the $n=1-5$ run extends electron distribution tail further out in the high velocity limit compared to the $n=1-2$ run. Also, current density function in velocity plotted in figure \ref{fig15fju}(c,d) manifests the same that current in the $n=1-5$ run is carried by comparatively higher energy electrons than the $n=1-2$ run. This characteristic prevails in multi-pass simulation for the X-wave as presented in detail in the preceding section. On other hand, this feature is absent from the O-wave multi-pass simulation result, eventhough all plasma and wave input parameters are kept exactly same with the X-wave simulation set-up. It is likely because the power deposition of O-wave on the 2nd harmonic is much weaker than that of the X-wave, what can't produce high velocity electron tail to resonate with higher harmonics. Also, it is unfolded in our simulation that such higher harmonics influence for the X-wave does actually occur above a threshold value of the EC wave's input power, indicating it to be a non-linear phenomenon dependant on the wave electric field. Mathematically, this nonlinearity comes from quadratic dependence of the QL diffusion co-efficient on the EC wave's electric field. The present article has advocated this fact as one of the plausible mechanism to explain the generation of energetic electrons from an initial low energy thermal electron population, and for the high current drive efficiency obtained in experiment. 

It is well known, from the earlier studies of particle acceleration via wave-particle interaction, charged particle may accelerate to very high velocity by means of ``overlapping of resonances" mechanism if the wave electric field exceeds stochasticity threshold~\cite{menyuk88}, and consequently, the motion of charged particle may lead to diffusion in velocity space~\cite{smith78}. Farina and co-authors~\cite{farina94} also reported that a local QL form of diffusion coefficient could accurately model this diffusive character of electron's motion. We think such physical mechanism may support our result of generation of energetic electron via multi-harmonic resonance that is obtained by numerically solving the FP equation with QL diffusion coefficients. However, this conjecture needs to be verified through further theoretical and numerical analysis, beyond the scope of present article though. 

The aforementioned physical process possibly explains high current drive efficiency for low plasma density experiments, but may not be exclusive in interpreting the plasma current measured at high density experiments. As shown in figure~\ref{fig11dt} of single-pass absorption simulation in a setup of $Z_{\rm eff}=4.0$, $I_p$ drops dramatically to low value upon three fold increasing of the core plasma density value $1\times10^{18}~m^{-3}$. In reality, as noticed in relatively higher density discharges on EXL-50~\cite{shi21}, plasma current falls with increasing line integrated density though, but not exhibiting a quick exponential drop like in simulation. However, we can't reach to a definite conclusion until full profiles of plasma temperature and density will be available from diagnostic measurement because figure~\ref{fig11dt} also indicates that higher temperature of electrons may keep the current level up to a necessarily high value. Whatsoever, this apparent discrepancy between experiment and simulation at high density scenario may be interpreted as follows:

\begin{itemize}
	\item The model implemented in CQL3D can not calculate current flowing along open field lines outside of the LCFS, which may although be a dominant fraction in experiment as estimated by a different physical model~\cite{ishida21}. Increasing density at low plasma temperature scenarios reduces $I_p$ to a substantially low value inside the LCFS, but it doesn't affect much the current outside  the LCFS, mostly carried by very high velocity electrons being less impacted upon collision with all background plasma species. Therefore, the total current measured in experiment may still remain quite large with contribution from the open field line region. 
	\item At density above the O-wave cut-off level, this scenario can possibly switch over from EC-waves to  electron Bernstein waves (EBW) being excited via mode conversion, and a good amount of current may be maintained by EBW inside the LCFS. 
\end{itemize}
Due to the limitation of diagnostics, at present, we don't have any idea about the ratio of current flowing in and out-side the LCFS. Further analysis will be carried out in this regard once required experimental data is available.  

Recent simulation studies on the QUEST ST~\cite{onchi21,kojima21} - an EC wave driven non-inductive plasma operation similar to EXL-50 - has also investigated the dominancy of energetic electron fraction over the bulk thermal electron population in absorbing EC-wave power, and predicted the usefulness of multi-harmonic on current drive by examining the relativistic ECR condition in high energy domain. In another modeling of non-inductive plasma current~\cite{ono20}, it was reported, in their terminology, minority hot electrons by the 2nd harmonic ECR power absorption could drive most of the current observed on QUEST. Our recent study of ray-tracing simulation on EXL-50 also confirmed that the single pass absorption ratio of X- to O-wave can be larger than $50\%$ with around $2\%$ energetic electrons of few hundreds keV energy~\cite{xie21}. In all of these studies, various energetic electron parameters - its distribution, temperature and density - are presumed and set through a mathematical function with some informations collected from experiments. However, in our present study, energetic electron tail has been formed through the FP equation solution in time starting with an initial low temperature ($<100eV$) Maxwellian distribution. Now, making assessment of how close our present simulation effort has been to match with the actual experiment on EXL-50, we may emphasize our result being consistent with the finding in experiment. However, there are some features, those may be influential in reality but could not be implemented in this work such as,
\begin{itemize}
\item No model is included to simulate current flowing along the open field line region. But, in reality, plasma may have a part of current outside the closed flux region as identified on basis of a multi-fluid modeling~\cite{ishida21}.

\item Radial diffusion of electrons is not considered in our modeling, that could expand the highly localized current density profile formed in simulation.

\item Orbital loss mechanism for fast electrons that has been found important in describing current on MAST~\cite{toit17} and LATE~\cite{maekawa12} is also not modeled here.
\end{itemize}
Nevertheless, our numerical study has revealed an effective mechanism of ``multi-harmonic interaction" that may explain high current drive efficiency for EXL-50.
\section{Summary}
Our results from a multi-pass ray-tracing simulation study demonstrate the effectiveness of multi-harmonic ECRs on plasma heating and current drive on EXL-50. Despite having a low density, low temperature plasma at start-up, this ST routinely achieves good efficiency in current (over 1 A/W) by sole ECRH drive. In simulation, by applying the Fokker-Plank, quasi-linear theory, same high efficiency in current has also been obtained in presence of multi-harmonic ECRs. This is because energetic electrons resonate with multiple higher harmonics via the relativistic Doppler shift. The effectiveness of higher harmonics ($n>2$) is found significant when the input wave power exceeds a threshold value, suggesting it to depend on the quadratic nonlinearity of wave electric field in the QL diffusion coefficient. Also found is that plasma current increases significantly in multi-pass absorptions compared to single-pass absorption in simulation, indicating the beneficial role of multi-reflection of EC waves inside the vacuum vessel. The current density profile emerged from this modeling is highly localized radially which differs from our observation of a broader profile in experiment, encompassing both the closed and open field regions~\cite{ishida21}. The absence of radial diffusion calculation of electrons across flux surfaces in our present model could possibly be a reason behind this outcome. It may also arise due to inability of CQL3D to track energetic electrons which eventually travel from closed to open field lines, and remain confined. In this regard, a complete model considering EC wave driven current in open field lines remains a necessity to develop in future. Further research will be carried out in experiment and modeling to better understand this mechanism of `multi-harmonic interaction' and improve the performance of EXL-50 operation. 
\section*{Acknowledgment}
Authors sincerely thank Dr.~Houyang Guo (ENN) for his assistance of careful reading the manuscript and commenting on it. DB highly appreciates Dr.~Bihe Deng's (ENN) interesting questions on the physics content, and acknowledges encouragement by Mr. Baoshan Yuan (ENN). This work was performed under the auspices and full support of the compact fusion project granted by the ENN group, China. Yu .V. Petrov was supported by the US Department of Energy under grant DE-FG02-04ER54744. 
\section*{References}
\bibliographystyle{iopart-num}
\bibliography{smbib}

\providecommand{\newblock}{}
\begin{thebibliography}{10}
\expandafter\ifx\csname url\endcsname\relax
  \def\url#1{{\tt #1}}\fi
\expandafter\ifx\csname urlprefix\endcsname\relax\def\urlprefix{URL }\fi
\providecommand{\eprint}[2][]{\url{#2}}

\bibitem{prater03}
Prater R 2004 {\em Phys. Plasmas\/} {\bf 11} 2349

\bibitem{karney78}
Karney C~F~F 1978 {\em Phys. Fluids\/} {\bf 21} 1584

\bibitem{luce99}
Luce T~C {\em et~al.\/} 1999 {\em Phys. Rev. Lett.\/} {\bf 83} 4550

\bibitem{harvey97}
Harvey R~W {\em et~al.\/} 1997 {\em Nucl. Fusion\/} {\bf 37} 69

\bibitem{menyuk88}
Menyuk C~R 1988 {\em Phys. Fluids\/} {\bf 31} 3768

\bibitem{harvey01}
Harvey R~W and Perkins F~W 2001 {\em Nucl. Fusion\/} {\bf 41} 1847

\bibitem{ikegami73}
Ikegami H {\em et~al.\/} 1973 {\em Nucl. Fusion\/} {\bf 13} 351

\bibitem{maekawa18a}
Maekawa T {\em et~al.\/} 2018 {\em Nucl. Fusion\/} {\bf 58} 016037

\bibitem{maekawa18b}
Maekawa T {\em et~al.\/} 2018 {\em Nucl. Fusion\/} {\bf 58} 106020

\bibitem{erckmann94}
Erckmann V and Gasparino U 1994 {\em Plasma Phys. Control. Fusion\/} {\bf 36}
  1869--1962

\bibitem{shi21}
Shi Y {\em et~al.\/} 2021 Solenoid-free current drive via ecrh in exl-50
  spherical torus plasmas arXiv:2104.14844

\bibitem{ishida21}
Ishida A {\em et~al.\/} 2021 {\em Phys. Plasmas\/} {\bf 28} 032503

\bibitem{maekawa05}
Maekawa T {\em et~al.\/} 2005 {\em Nucl. Fusion\/} {\bf 45} 1439--1445

\bibitem{idei17}
Idei H {\em et~al.\/} 2017 {\em Nucl. Fusion\/} {\bf 57} 126045

\bibitem{onchi21}
Onchi T {\em et~al.\/} 2021 {\em Phys. Plasmas\/} {\bf 28} 022505

\bibitem{shevchenko10}
Shevchenko V~F {\em et~al.\/} 2010 {\em Nucl. Fusion\/} {\bf 50} 022004

\bibitem{shevchenko15}
Shevchenko V~F {\em et~al.\/} 2015 {\em EPJ Web of Conf.\/} {\bf 87} 02007

\bibitem{harvey92}
Harvey R~W and McCoy M~G 1992 The cql3d fokker-plank code proc. of iaea tcm
  montreal, June 15-18

\bibitem{harvey94}
Harvey R~W 1994 {\em Bull. Am. Phys. Soc.\/} {\bf 39} 1626

\bibitem{li21}
Li S~J {\em et~al.\/} 2021 {\em JINST\/} {\bf 16}

\bibitem{cheng21}
Cheng S~K {\em et~al.\/} 2021 {\em Rev. Sci. Instrum.\/} {\bf 92} 043513

\bibitem{uchida10}
Uchida M {\em et~al.\/} 2010 {\em Phys. Rev. Lett.\/} {\bf 104} 065001

\bibitem{maekawa12}
Maekawa T {\em et~al.\/} 2012 {\em Nucl. Fusion\/} {\bf 52} 083008

\bibitem{kerbel85}
Kerbel G~D and McCoy M~G 1985 {\em Phys. Fluids\/} {\bf 28} 3629

\bibitem{braams89}
Braams B~J and Karney C~F~F 1989 {\em Phys. Fluids\/}  1355

\bibitem{stix}
Stix T~H 1992 {\em Waves in Plasmas\/} (New york: AIP)

\bibitem{cql3d}
Harvey R~W and McCoy M~G 1992 The cql3d fokker-plank code (\textit{Preprint}
  \eprint{available at http://compxco.com/cql3dmanual.pdf})

\bibitem{prater08}
Prater R {\em et~al.\/} 2008 {\em Nucl. Fusion\/} {\bf 48} 035006

\bibitem{harvey02}
Harvey R~W {\em et~al.\/} 2002 {\em Phys. Rev. Lett.\/} {\bf 88} 205001

\bibitem{zheng18}
Zheng P {\em et~al.\/} 2018 {\em Nucl. Fusion\/} {\bf 58} 036010

\bibitem{petty02}
Petty C~C {\em et~al.\/} 2002 {\em Nucl. Fusion\/} {\bf 42} 1366

\bibitem{smith78}
Smith G~R and Kaufman A~N 1978 {\em Phys. Fluids\/} {\bf 21} 2230

\bibitem{farina94}
Farina D {\em et~al.\/} 1994 {\em Phys. Plasmas\/} {\bf 1} 1871

\bibitem{kojima21}
Kojima S {\em et~al.\/} 2021 {\em Plasma Phys. Control. Fusion\/} {\bf 63}
  105002

\bibitem{ono20}
Ono M {\em et~al.\/} 2020 {\em AIP Conf. Proc.\/} {\bf 2254} 090001

\bibitem{xie21}
Xie H~S {\em et~al.\/} 2021 Boray: An axisymmetric ray tracing code supports
  both closed and open field lines plasma arXiv:2105.12014

\bibitem{toit17}
du~Toit E~J {\em et~al.\/} 2017 {\em EPJ Web of Conf.\/} {\bf 147} 01002

\end{thebibliography}
\end{document}